\documentclass[aps,notitlepage,twocolumn,showpacs,preprintnumbers,floatfix,tightenlines, longbibliography, nofootinbib, superscriptaddress]{revtex4-2}

\usepackage{graphicx} 
\usepackage{hyperref}
\usepackage{amsmath,bm}
\usepackage[capitalise]{cleveref}
\usepackage{color}
\usepackage{comment}
\usepackage{soul} 

\def\k{\bm{k}}
\def\x{\bm{x}}
\def\rmd{{\rm d}}

\def\q{\bm{q}}
\def\p{\bm{p}}
\newcommand{\vb}[1]{\mathbf{#1}}
\newcommand{\dd}[2][]{\mathrm d^{#1}{#2}\,} 
\newcommand{\GretHTL}{G^{\mathrm{isoHTL}}}
\newcommand{\jet}{{\mathrm{multi}}}
\newcommand{\IM}{\mathrm{Im\,}}
\newcommand{\RE}{\mathrm{Re\,}}

\def\am#1{{\color{green}\bf[#1]\text{$_{\rm AM}$}}} 

\def\flo#1{{\color{cyan}\bf[#1]\text{$_{\rm FL}$}}} 
\def\kb#1{{\color{red}\bf[#1]\text{$_{\rm KB}$}}} 

\begin{document}
\title{Minijet thermalization and jet transport coefficients in QCD kinetic theory}

\author{Kirill Boguslavski}
\email{kirill.boguslavski@subatech.in2p3.fr}
\affiliation{SUBATECH UMR 6457 (IMT Atlantique, Université de Nantes, IN2P3/CNRS), 4 rue Alfred Kastler, 44307 Nantes, France}
\affiliation{Institute of Theoretical Physics, TU Wien, Wiedner Hauptstraße 8-10, 1040 Vienna, Austria}
\author{Florian Lindenbauer}
\email{flindenb@mit.edu}
\affiliation{MIT Center for Theoretical Physics – a Leinweber Institute, Massachusetts Institute of Technology, Cambridge, MA 02139, USA}
\affiliation{Institute of Theoretical Physics, TU Wien, Wiedner Hauptstraße 8-10, 1040 Vienna, Austria}
\author{Aleksas Mazeliauskas}
\email{a.mazeliauskas@thphys.uni-heidelberg.de}
\affiliation{Institute for Theoretical Physics, University of Heidelberg, Philosophenweg 16, 69120 Heidelberg, Germany}
\author{Adam Takacs}
\email{takacs@thphys.uni-heidelberg.de
}
\affiliation{Institute for Theoretical Physics, University of Heidelberg, Philosophenweg 16, 69120 Heidelberg, Germany}
\author{Fabian Zhou}
\email{zhou@thphys.uni-heidelberg.de}
\affiliation{Institute for Theoretical Physics, University of Heidelberg, Philosophenweg 16, 69120 Heidelberg, Germany}

\preprint{MIT-CTP/5948}

\begin{abstract}
We apply weakly coupled QCD kinetic theory to investigate the thermalization of high-momentum on-shell partons (minijets) in a Quark-Gluon Plasma (QGP). Our approach incorporates isotropic hard thermal loop screening to model soft quark and gluon exchanges, allowing us to verify consistency with established analytic results of jet transport coefficients. 
We perform kinetic simulations of minijets propagating through a thermal gluon plasma, incorporating both collinear radiation and elastic scatterings. The resulting evolution is compared to predictions from jet transport coefficients, including the longitudinal and transverse jet-quenching parameters $\hat q$, energy loss, and the drag coefficient. We find that standard definitions of jet transport coefficients neglect the contributions from recoiling
medium particles. Including these contributions restores consistency with the kinetic evolution. Finally, we show that the minijet thermalization time scales remarkably well with $\hat{q}$ and we produce a phenomenological estimate of the minijet quenching time in heavy-ion collisions.
\end{abstract}

\maketitle

\section{Introduction}

The experimental and theoretical study of relativistic nuclear collisions has established a detailed picture of how the Quark–Gluon Plasma (QGP) forms under extreme conditions of temperature and density~\cite{Busza:2018rrf,Schlichting:2019abc,Berges:2020fwq,Cao:2024pxc}. Its properties are explored through a wide range of observables, including low- and high-momentum hadrons, electromagnetic radiation, and heavy-flavor probes. Among the most prominent phenomenological signatures of QGP formation is the energy loss experienced by high-momentum partons traversing the medium. This process leads to a suppression of high-momentum hadrons and reconstructed jets (jet quenching) relative to an equivalent proton–proton baseline~\cite{Casalderrey-Solana:2007knd,dEnterria:2009xfs,Armesto:2015ioy,Connors:2017ptx,Cunqueiro:2021wls}.
Hard scatterings initially produce highly virtual partons that undergo rapid vacuum-like splittings. As they lose virtuality and become on-shell, further splittings occur only through medium interactions. We refer to these on-shell partons as minijets to distinguish them from high-virtuality partons in vacuum showers.
A detailed theoretical understanding of how energetic partons interact with the evolving medium remains an outstanding problem. 
Here we study how the energy and momentum transport of high-momentum partons relate to the evolution and thermalization of minijets in a kinetic framework.

Jets traversing the Quark–Gluon Plasma are commonly characterized by transport coefficients such as the transverse/longitudinal momentum broadening coefficient $\hat q$, longitudinal drag $\eta$, and conversion rate $\hat \Gamma$~\cite{Ghiglieri:2015zma}. These coefficients have been studied extensively within different theoretical frameworks: leading and next-to-leading order QCD kinetic theory~\cite{Arnold:2002zm,Ghiglieri:2015ala}, non-perturbative Euclidean lattice QCD~\cite{Panero:2013pla,Moore:2019lgw}, light-front QCD Hamiltonian approach~\cite{Li:2023jeh}, glasma simulations~\cite{Ipp:2020nfu, Avramescu:2023qvv, Carrington:2021dvw}, 
holography~\cite{Liu:2006ug,Casalderrey-Solana:2011dxg}, in inhomogeneous and flowing media \cite{Barata:2022krd, Kuzmin:2023hko} and other frameworks 
\cite{Uphoff:2014cba, Grishmanovskii:2024gag, He:2015pra}. 
On the phenomenological side, jet transport properties have also been constrained using collider data and end-to-end event generators~\cite{JET:2013cls,Paquet:2023rfd,JETSCAPE:2021ehl,Karmakar:2023ity,JETSCAPE:2024cqe}. 

While transport coefficients provide valuable insight, they have intrinsic limitations. By construction, they describe an idealized single energetic parton propagating through a medium, assuming a large-scale separation between the jet energy and the medium temperature. A more realistic description of jet quenching requires frameworks that can follow the dynamics of energetic probes from their initialization to eventual thermalization. 
Within the framework of QCD kinetic theory simulations, the minijet thermalization studies have been done in Refs.~\cite{Cao:2020wlm,Schlichting:2020lef,Mehtar-Tani:2022zwf,Boguslavski:2023alu,Boguslavski:2023waw,Boguslavski:2023jvg,Zhou:2024ysb}. Such an approach also allows one to investigate jet–medium interactions in out-of-equilibrium conditions, as realized in the early stages of realistic nuclear collisions.

In this work, we employ the leading-order QCD effective kinetic theory (EKT) to directly compare the traditional transport-coefficient description of jet quenching with the full microscopic evolution of on-shell jets in a medium. For simplicity, we restrict the study to the Yang–Mills sector, where gluonic perturbations propagate through a thermal gluon plasma.
The EKT framework, which has been successfully applied to study the chemical, kinetic, and hydrodynamic thermalization of the QGP as well as electromagnetic probes~\cite{Baier:2000sb,Schlichting:2019abc,Berges:2020fwq,Kurkela:2014tea,Kurkela:2015qoa,Kurkela:2018oqw,Kurkela:2018xxd,Du:2020dvp,Du:2020zqg}, also contains the essential physics of high-momentum parton energy loss~\cite{Ghiglieri:2015zma}. For recent applications to minijet thermalization in static and expanding plasmas see Refs.~\cite{Schlichting:2020lef,Mehtar-Tani:2022zwf,Boguslavski:2023alu,Boguslavski:2023waw,Zhou:2024ysb}.
In this work, we employ an isotropic HTL screening prescription for elastic scatterings~\cite{Boguslavski:2024kbd} and demonstrate consistency with analytic results for jet transport coefficients in the infinite jet-energy limit\footnote{Although the main text focuses on the Yang–Mills plasma, in \cref{app:HTL_Screening} we derive the isotropic HTL screening prescription for soft quark exchanges and compute the corresponding quark conversion rate.}.
In contrast, for finite minijet momentum, we show that traditionally neglected effects, such as medium recoil and radiative corrections, play an important role in defining transport coefficients.
We formulate a cutoff-dependent definition of transport coefficients by including contributions from all particles with momenta $p > \Lambda_{\text{cutoff}}$ and analyze the resulting cutoff dependence.
Our analysis clarifies the connection between transport coefficients and full kinetic theory evolution of jet quenching. Finally, we follow the minijet evolution up to thermalization and extract the corresponding thermalization time, which scales remarkably well with $\hat q$. This relation provides a compact parametrization for minijet thermalization in the medium.

The paper is organized as follows. In \cref{sec:EKTintro} we briefly describe the EKT framework and the initial conditions of the EKT evolution. In particular, we comment on the implementation of isotropic HTL screening used in this work. \Cref{sec:jetcoef} introduces the definitions of transport coefficients in kinetic theory and highlights the approximations made in the traditional definitions. We then make a numerical comparison of transport coefficients extracted from the EKT evolution and those from direct computations for different low-momentum cutoffs. In \cref{sec:resultsthermalization} we study minijet thermalization and show the scaling of the thermalization time with transport properties. Our conclusions are given in \cref{sec:conclusions}. For completeness, our paper included a number of appendices. In \cref{App:EKT}, we present explicit expressions for the collision kernels. In \cref{app:HTL_Screening}, we detail the implementation of isotropic HTL screening for soft gluon and quark exchanges and validate the computed transport coefficients in the infinite jet-momentum limit against known analytic results.
In \cref{app:Expectation_values_OE} we define the expectation value of a general observable in the single collision limit and in \cref{eq:numerics} we provide the corresponding numerical implementation details.
Finally, we demonstrate the convergence of transport coefficient extraction from the EKT evolution in \cref{app:convergence}.

\section{QCD kinetic theory setup}\label{sec:EKTintro}
The high-temperature QGP can be described as a weakly interacting gas of quarks and gluons. The particles are then characterized by phase-space distribution functions $f_a(t, \x, \p)=\frac{(2\pi)^3}{\nu_a}\frac{\mathrm{d} N_a}{\mathrm{d}^3\x \mathrm{d}^3 \p}$, where $a$ stands for the particle species and $\nu_a$ is its number of degrees of freedom, $16$ for gluons and $6$ for each flavor of quarks and antiquarks. The distribution functions obey the relativistic Boltzmann equation
\begin{equation}\label{eq:Boltz}
    \left[\partial_t + \frac{\p}{p}\cdot \nabla_{\x}\right]f_a(t,\x,\p) = C_a[\{f\}](t,\x,\p)\,,
\end{equation}
where $C_a$ encodes the collision processes.
The relevant collision processes were systemized by Arnold, Moore and Yaffe (AMY) \cite{Arnold:2002zm} using the hard thermal loop framework at leading order in the QCD coupling $\lambda=g^2N_c$.
They consist of elastic $2\leftrightarrow 2$ scatterings and collinear $1\leftrightarrow 2$ inelastic collisions for
massless particles $P^0 =|\p|=p$ at sufficiently high momentum ($p\gg g T$).

The elastic collision kernel is given by a multi-dimensional momentum integral
\begin{equation} \label{eq:C22}
    \begin{split}
        &C_a^{2\leftrightarrow 2}[\{f\}] = \frac 1{4p\nu_a} \sum_{bcd} \int d\Omega^{2\leftrightarrow2}
        \left|{\cal M}^{ab}_{cd}\right|^2\mathcal{F}^{2\leftrightarrow2} \,,
    \end{split}
\end{equation}
where we used the shorthand notation for the phase space integral
\begin{equation}
    \begin{split}
        \int d\Omega^{2\leftrightarrow2}&= \int_{p_2p_3p_4} (2\pi)^4\delta^{4}(P+P_2-P_3-P_4)\,.
    \end{split}
\end{equation}
with $\int_{\p}=\int\frac{\rmd^3\p}{(2\pi)^3}$, and $\int_{p}=\int_{\p}\frac{1}{2p}$. Here, the delta function imposes the 4-momentum conservation. The statistical gain-loss factor reads
\begin{equation}\label{eq:stat22}
    \begin{split}
        \mathcal{F}^{2\leftrightarrow2} &= f_c(\bm p_3)f_d(\bm p_4)(1\pm f_a(\bm p))(1\pm f_b(\bm p_2))\\
        &-f_a(\bm p)f_b(\bm p_2)(1\pm f_c(\bm p_3))(1\pm f_d(\bm p_4))\,,
    \end{split}
\end{equation}
with $+$ for bosons and $-$ for fermions, and all $f$ evaluated at the same $(t,\x)$. 

The $\left|{\cal M}^{ab}_{cd}\right|^2$ is the spin-, and color-summed hard thermal loop (HTL) matrix element (see \cref{app:HTL_Screening}). In the vacuum, it has a pole in the small-momentum exchange limit, while at finite temperature, due to screening, it is regulated by in-medium masses. In previous works~\cite{AbraaoYork:2014hbk, Kurkela:2014tea, Kurkela:2015qoa, Kurkela:2018oqw, Kurkela:2018xxd, Kurkela:2021ctp, Du:2020dvp, Du:2020zqg, Kurkela:2018vqr, Kurkela:2018wud}, these matrix elements were approximated by simple Debye-like screened vacuum matrix elements including a soft-momentum regulator
(Debye-like prescription), whose parameter $\xi$ was chosen to reproduce longitudinal gluon momentum broadening and the gluon-to-quark conversion rate in equilibrium.

Recently, the more complete isotropic HTL screening prescription was used in the Boltzmann equation in Refs.~\cite{Mehtar-Tani:2022zwf,Boguslavski:2023waw,Boguslavski:2024kbd} instead of its approximated Debye-like prescription, and differences regarding the shear viscosity and the pressure anisotropy were identified. 
Here, we extend previous work by demonstrating that the isotropic HTL screening prescription simultaneously reproduces the analytical results for longitudinal and transverse momentum diffusion in the infinite jet-momentum limit.
We further develop the isotropic HTL screening for soft quark exchanges and show that it reproduces the gluon-to-quark conversion rate in the same limit.
Since the remainder of this paper focuses on minijets of finite momentum in a gluonic plasma, we delegate the detailed description of the HTL screening implementation and its validation against results in the literature to \cref{app:HTL_Screening}.

The inelastic collision kernel is given by
\begin{equation}\label{eq:C12}
    \begin{split}
         C_a^{1\leftrightarrow 2}[\{f\}] = \frac{(2\pi)^3}{2|\p|^2\nu_a}\sum_{bc}&\int_0^\infty dp'dk'\delta(p-p'-k')\\
       &  \times \gamma^{a}_{bc}\mathcal F^{1\leftrightarrow 2}(\p;p'\hat\p,k'\hat\p)\\
        -\frac{(2\pi)^3}{|\p|^2\nu_a}\sum_{bc}&\int_0^\infty dp'dk'\delta(p+k'-p')\\
        &\times \gamma^c_{ab}\mathcal F^{1\leftrightarrow 2}(p'\hat\p;\p,k'\hat\p)\,,
    \end{split}
\end{equation}
where the statistical factor reads
\begin{equation}\label{eq:stat12}
    \begin{split}
        \mathcal F^{1\leftrightarrow2}(\p_1;\p_2,\p_3) &= f_b(\p_2)f_c(\p_3)(1\pm f_a(\p_1))\\
        &-f_a(\p_1)(1\pm f_b(\p_2))(1\pm f_c(\p_3))\,,
    \end{split}
\end{equation}
and is evaluated at the same $(t,\x)$. Here, $\gamma^a_{bc}$ is the rate of collinear splittings $a\to bc$, and is obtained by resumming multiple interactions with the plasma \cite{Arnold:2002zm}, see \cref{App:EKT}.
In practice for numerical efficiency, we interpolate between the Bethe-Heitler and Landau-Pomeranchuk-Migdal limits, which reproduces well the results of
the linear integral equation, which sums up multiple interactions, see Appendix B of Ref.~\cite{Kurkela:2022qhn}.

In this work we will focus on a thermal Yang-Mills plasma at constant temperature $T$, and where its thermal distribution $n$ (Bose-Einstein for gluons) is perturbed by a ``minijet" at some energy scale $E$ (but homogeneous in space). To follow the evolution of the minijet, we employ the linearized Boltzmann equation, which is obtained by substituting $f=n+\delta f$ in \cref{eq:Boltz} and keeping only terms linear in $\delta f$. The linearized collision kernels have been discussed in previous works and we do not repeat the lengthy expressions here~\cite{Zhou:2024ysb}. We note that the collision matrix element $\left|{\cal M}^{ab}_{cd}\right|^2$ and splitting rate $\gamma^a_{bc}$ depend implicitly on the distribution function $f$. Therefore, generally, the linearized collision kernels should also contain linear corrections to $\left|{\cal M}^{ab}_{cd}\right|^2$ and $\gamma^a_{bc}$ (see \cite{Zhou:2024ysb}). However, in thermal equilibrium, these corrections are multiplied by the statistical factors \cref{eq:stat22,eq:stat12} that vanish identically. Thus, in the considered case of thermal background, the linearized collision kernels contain only those linearized terms that originate from the statistical factors.

\section{Jet transport coefficients from the Boltzmann equation}\label{sec:jetcoef}

In this section, we introduce the EKT linearization framework and minijet dynamics, and we derive their relationship with traditional definitions and computations of jet transport coefficients. Finally, we make a numerical comparison between transport properties extracted from the linearized EKT evolution and from a direct evaluation.

\subsection{Linearized Boltzmann equation and observables}

The linearized Boltzmann equation for the minijet as a small perturbation around the homogeneous medium 
($f=n+\delta f$) 
is given by
\begin{equation}
    \label{eq:LB}
    \partial_t \delta f_i(t,\bm p) = \delta C_i[\{n,\delta f\}]\,,
\end{equation}
where $\delta C_i[\{n,\delta f\}]$ is the linearized collision term. It contain elastic and inelastic scatterings $\delta C_i=\delta C_i^{2\leftrightarrow2}+\delta C_i^{1\leftrightarrow2}$ that originate from \cref{eq:C22} and \cref{eq:C12} by linearization as detailed in \cref{App:EKT}.

As we consider a non-expanding and homogeneous gluonic medium in thermal equilibrium, the background medium satisfies the Bose-Einstein distribution $n(p)=1/(e^{p/T}- 1)$.
The initial condition for the minijet is, however, taken as a perturbation at momentum $\bm p_0$. For analytical calculations, we employ a Delta function, 
\begin{equation}
    \label{eq:init_delta}
    \delta f_i(t{=}t_0, \p)=\epsilon\, T^3\times (2\pi)^3\delta_{ii_0} \delta^{(3)}(\bm p-\bm p_0)\,,
\end{equation}
with initial species index $i_0$. 
As the distribution function must be dimensionless, we have included a $T^3$ factor multiplied by an arbitrary dimensionless parameter $\epsilon$, where we assume that  $\epsilon \ll 1$ that justifies the linearization. 
Note that for numerical simulations of \cref{eq:LB}, we will instead use  a narrow Gaussian function that will be introduced later.

We define jet observables through the expectation value
\begin{equation}\label{eq:exp_value}
    \begin{split}
        \langle \mathcal O(\p;i)\rangle 
        = \frac{\sum_i\nu_i\int_{\bm p}\mathcal O(\p;i)\,\delta f_i(t,\bm p)}{\sum_i\nu_i\int_{\bm p}\delta f_i(t{=}t_0, \p)}\,,
    \end{split}
\end{equation}
where the denominator is evaluated for the initial distribution at $t=t_0$.
Note that this definition allows for multiple scatterings and thus a nontrivial evolution of the observables due to the dynamics of the minijet $\delta f_i(t,\bm p)$, even for a non-evolving medium, such as in our case.
This allows us to study jet thermalization due to multiple elastic and inelastic scatterings with the medium, as well as transport coefficients by using a single-hit picture, which we can then compare to results in the literature.

Typical time-dependent observables that we consider are the average transverse and longitudinal momenta squared of the jet, which quantify momentum broadening (see also \Cref{sec:resultsthermalization})
\begin{equation}
    \langle (\Delta\bm p_\perp)^2\rangle, \quad \langle (\Delta\bm p_{\parallel})^2\rangle, \quad \langle (\bm p_{\parallel})^2\rangle\,.\label{eq:p_obs}
\end{equation}
Here, we have decomposed the momentum $\bm p$ into its part parallel and transverse to the propagation of the minijet initiator, $\bm p=\bm p_{\parallel}+\bm p_\perp$, where $\bm p\cdot\hat\p_0=p_{\parallel}$, and $\bm p_\perp\cdot\hat\p_0=0$. The transverse momentum with respect to $\bm p_0$ is then $\Delta\bm p_\perp=\bm p_\perp=(I-\frac{\bm p_0\otimes\bm p_0}{p_0^2})\cdot\bm p$, while we distinguish relative $\Delta\bm p_{\parallel}=\bm p_{\parallel}-\bm p_0=(\frac{\bm p\cdot \bm p_0}{p^2_0}-1)\bm p_0$ and absolute longitudinal broadening.

\subsection{Collisional expansion and transport coefficients}

The linearized QCD kinetic theory can be used to redefine jet transport coefficients as single-hit observables and numerically compute them. To achieve this,
we expand the jet distribution in the number of collisions, such that 
\begin{equation} \label{eq:opacity_expansion}
    \delta f_i=\delta f_i^{(0)}+\delta f_i^{(1)}+\dots\,. 
\end{equation}
The first two orders are given by
\begin{equation} \label{eq:single-hit}
    \begin{split}
        \partial_t\delta f_i^{(0)}&=0\,, \\
        \partial_t\delta f_i^{(1)}&=\delta C_i[\{n,\delta f^{(0)}\}]\,,
    \end{split}
\end{equation}
where $\delta f^{(0)}$ is the free propagation of the initial perturbation, which is constant in the non-expanding thermal case $\delta f_i^{(0)}(t,\bm p)=\delta f_i(t{=}t_0, \p)$. Then, $\delta f^{(1)}$ describes the distribution of particles after one collision (including elastic and inelastic processes). 
Similar to the zeroth order, $\partial_t\delta f^{(1)}$ is time independent for our spacetime independent distributions $n$ and $\delta f^{(0)}$ given by \cref{eq:init_delta}.

The expansion \eqref{eq:opacity_expansion} has been previously referred to as ``opacity expansion'' with the first orders \eqref{eq:single-hit} forming the ``single-hit'' approximation~\cite{Kurkela:2018qeb, Kurkela:2021ctp, Ambrus:2022qya}. We note, however, that strictly speaking, a single hit of the medium refers to sufficiently {\em hard} medium particles with momenta $k \sim T$. In our approach, the linearized collision kernels go slightly beyond it as they contain HTL-resummed propagators, which is consistent with the kinetic evolution of the medium.
Understanding jet observables within the EKT framework and in its opacity expansion aligns well with recent theoretical developments in jet quenching~\cite{Mehtar-Tani:2019tvy,Barata:2020rdn,Barata:2021wuf,Isaksen:2022pkj} and with the isolation of the single/multiple scattering regimes in phenomenological applications~\cite{Caucal:2021cfb,Cunqueiro:2023vxl,Andres:2023xwr,Barata:2023bhh} and experimentally~\cite{ALICE:2024fip,CMS-PAS-HIN-24-016}.

Using the single-hit formulation \eqref{eq:single-hit} of the expectation values \eqref{eq:exp_value} indicated as $\langle \cdot \rangle^{(1)}$, we can conveniently define different transport coefficients and express them in terms of our collision kernels%
\footnote{
Note that for our initial conditions, the transport coefficients defined in the single-hit expansion are time-independent.
}
\begin{align}
    \hat q_\perp &= \partial_t\langle (\Delta\bm p_\perp)^2\rangle^{(1)}\nonumber \\
    &=\sum_i\frac{\nu_i}{\nu_{i_0} \epsilon\, T^3}\int_{\bm p}(\Delta \bm p_{\perp})^2\delta C_i[\{n,\delta f^{(0)}\}]\,,\label{eq:LB_qhat} \\
    \hat q_{\parallel} &= \partial_t\langle (\Delta \bm p_{\parallel})^2\rangle^{(1)} \nonumber \\
    &= \sum_i\frac{\nu_i}{\nu_{i_0} \epsilon\, T^3}\int_{\bm p}(\Delta \bm p_{\parallel})^2\delta C_i[\{n,\delta f^{(0)}\}]\,,\label{eq:LB_qhat_parallel} \\
    \hat q_L&=\partial_t\langle \bm p_{\parallel}^2\rangle^{(1)} \nonumber \\
    &= \sum_i\frac{\nu_i}{\nu_{i_0} \epsilon\, T^3}\int_{\bm p}\bm p^2_{\parallel}\, \delta C_i[\{n,\delta f^{(0)}\}]\,, \label{eq:LB_qhat_L}
    \end{align}
    \begin{align}
    \eta_D &= -\frac{1}{\langle p_{\parallel}\rangle^{(0)}}\partial_t\langle p_{\parallel}\rangle^{(1)} \nonumber \\
    &= -\frac{1}{\langle p_{\parallel}\,\rangle^{(0)}}\sum_i\frac{\nu_i}{\nu_{i_0} \epsilon\, T^3}\int_{\bm p}|\bm p_{\parallel}|\,\delta C_i[\{n,\delta f^{(0)}\}]\,,\label{eq:LB_drag} \\
    \hat e &= \partial_t\langle |\p|\rangle^{(1)} \nonumber \\
    &= \sum_i\frac{\nu_i}{\nu_{i_0} \epsilon\, T^3}\int_{\bm p} |\p|\,\delta C_i[\{n,\delta f^{(0)}\}]\,,\label{eq:LB_e_loss} \\
    \hat\Gamma_{i_0} &= \partial_t\langle 1 \rangle_{i\neq i_0}^{(1)} \nonumber \\
    &= \sum_{i\neq i_0}\frac{\nu_i}{\nu_{i_0} \epsilon\, T^3}\int_{\bm p}\delta C_i[\{n,\delta f^{(0)}\}]\,.\label{eq:LB_Gamma}
\end{align}
Thus, we have defined the jet quenching parameter $\hat q_\perp$ and the longitudinal jet quenching parameter $\hat q_{\parallel}$ ($\hat q_{L}$) through transverse and relative (absolute) longitudinal momentum broadening of the initial perturbation. Following Ref.~\cite{Ghiglieri:2015ala}, we have also defined the longitudinal drag coefficient $\eta_D$ and the conversion rate $\hat\Gamma_{i_0}$. 
Note that the collisional energy loss coefficient $\hat e$ is closely related but not identical to the drag coefficient $\eta_D$, because most (but not all) of the energy loss of the jet will originate from the change in the longitudinal momentum. In leading-order kinetic theory, where inelastic processes are strictly collinear, the contribution to $\eta_D$ and $\hat e$ from these processes is identical (up to the factor $\langle p_{\parallel}\rangle$). This is only approximately true for the elastic processes. Thus, the only differences between $\hat e$ and $-\langle p_{\parallel}\rangle\eta_D$ must be due to elastic collisions.

The general expressions for the expectation values \cref{eq:exp_value} in the single-hit approximation are detailed in \cref{app:Expectation_values_OE} and have been partially used in linearized QCD kinetic theory frameworks before
\cite{Arnold:2000dr,Arnold:2003zc,Arnold:2008vd,York:2008rr,Ghiglieri:2018dgf,Kurkela:2021ctp,Tornkvist:2023kan}. 
Since the elastic matrix element and the splitting rate entering the collision kernels also depend on the medium, they include terms that additionally contribute to the linearized expressions. However, they are multiplied by the loss-minus-gain structures \eqref{eq:stat22} and \eqref{eq:stat12}, which vanish in thermal equilibrium identically, as in our case. We emphasize that, in contrast, in an evolving medium (cf.~\cite{Boguslavski:2023alu}), this needs to be taken into account, which modifies the usual expressions for transport coefficients.

In the following, we provide simple approximate expressions for the transport coefficients to illustrate their main contributions and properties.
For $\hat q_\perp$, 
inelastic collisions $\delta C^{1\leftrightarrow2}$ do not change the transverse momentum within a single scattering event, and one only needs to take elastic scatterings $\delta C^{2\leftrightarrow2}$ into account. 
Aligning $\p_0$ with the $z$-axis ($p_{0,\perp}=0$) leads to $\p_\perp^2=(p\sin\theta)^2$, where $\theta$ is the polar angle, and assuming $p_0\gg T$ yields the approximate form
\begin{equation}\label{eq:qhat_OE}
    \begin{split}
        \hat q_\perp & \approx \frac{1}{2p_0}\sum_{bcd}\int_{k,p'>k'}(2\pi)^4\delta^4(p_0+k-p'-k')\\
        &\times|\mathcal M^{i_0b}_{cd}(\p_0,\k;\p'\k')|^2  \left[n_b(k)(1\pm n_d(k'))\right] \\
        & \times [{\p'_\perp}^2+{\k'_\perp}^2-\k_\perp^2]\,.
    \end{split}
\end{equation}
\Cref{eq:qhat_OE} should provide a good approximation of $\hat q_\perp$ 
for sufficiently short path lengths relative to the mean free paths of elastic scatterings and radiation. 
Note that \cref{eq:qhat_OE} slightly differs from the standard definition of $\hat q$~\cite{Aurenche:2002pd, Arnold:2008vd, Ghiglieri:2015ala, Boguslavski:2023waw}
\begin{equation}\label{eq:qhat_stand}
    \begin{split}
        \hat q_\perp^{\text{stand}} & \approx \frac{1}{2p_0}\sum_{bcd}\int_{k,p'>k'}{\p'_\perp}^2 (2\pi)^4\delta^4(p_0+k-p'-k')\\
        &\times|\mathcal M^{i_0b}_{cd}(\p_0,\k;\p'\k')|^2  \left[n_b(k)(1\pm n_d(k'))\right],
    \end{split}
\end{equation}
where the additional terms $\k'_\perp{}^2-\k_\perp^2$ are missing,
as discussed in the next subsection.

In contrast, for longitudinal momentum broadening, radiative processes dominate over elastic scatterings, and the expression for the longitudinal jet quenching parameter could be approximated by
\begin{equation}
    \begin{split}
        \hat q_{\parallel}&\approx\frac{(2\pi)^3}{p_0^2}\sum_{ac}\int_0^\infty \rmd k\,\left[{k}^2-(k-p_0)^2\right]\\
        &\quad\times n_c(k)\gamma^{a}_{i_0c}((p_0+k)\hat\p_0;p_0\hat\p_0,k\hat\p_0) \\
        &+\frac{(2\pi)^3}{p_0^2}\sum_{ab}\int_0^{p_0/2} \rmd k\,\left[k^2+(k-p_0)^2\right]\\
        &\quad\times(1\pm n_b(k))\gamma^{i_0}_{ab}(p_0\hat\p_0;(p_0-k)\hat\p_0,k\hat\p_0)\,.
    \end{split}
\end{equation}

Similar approximate formulae can be derived for other transport coefficients, where in general both elastic and inelastic contributions have to be taken into account. The impact of each of the two contributions will be assessed in the following subsections. Moreover, the approximate analytical expressions of the transport coefficients can be compared to the standard formulations of the transport coefficients encountered in the literature, as mentioned above. We will provide comparisons analytically in \cref{sec:analyt_comp_transport} and numerically in \cref{sec:resultstransport} using the full expressions given by \cref{eq:LB_qhat,eq:LB_qhat_L,eq:LB_qhat_parallel,eq:LB_drag,eq:LB_e_loss,eq:LB_Gamma}.

\subsection{Difference between single-hit collision and HTL transport coefficients}
\label{sec:analyt_comp_transport}
Our result in Eq.~\eqref{eq:qhat_OE} for the jet quenching parameter is very similar to previous determinations of $\hat q_\perp$ in the high-temperature effective kinetic theory framework \eqref{eq:qhat_stand} (e.g., see \cite{Aurenche:2002pd, Arnold:2008vd, Ghiglieri:2015ala, Boguslavski:2023waw}).
The only difference is the additional term ${\k'_\perp}^2-\k_\perp^2$ in Eq.~\eqref{eq:qhat_OE}, which can be attributed to the \textit{broadening of the recoiling medium} and which was omitted in previous works. In particular, this term can be associated with the \textit{wake} that the jet leaves when traversing the medium. 
However, due to the consistency of the LO kinetic approach, this term should not be neglected when considering the linearized Boltzmann equation. Furthermore, previous studies focused mainly on elastic collisions $\delta C^{2\leftrightarrow2}$ when evaluating transport coefficients, while 
coefficients like $\hat q_{\parallel/L}$, $\eta_D$, or $\hat \Gamma_{i_0}$ are dominated by \textit{radiative processes} captured in $\delta C^{1\leftrightarrow2}$. 
In particular, traditionally, radiative contributions have been treated as higher-order corrections, while here we treat them within the single-hit expansion framework and show that they are non-negligible. Similar efforts have been recently done in Refs.~\cite{Liou:2013qya,Blaizot:2014bha,Blaizot:2019muz,Arnold:2021mow,Arnold:2021pin} identifying logarithmically enhanced corrections to $\hat q_\perp$. Beyond the single-hit limit, the Boltzmann equation re-iterates these radiative corrections, which has become of recent interest~\cite{Caucal:2021lgf,Caucal:2022fhc,Caucal:2022mpp,Arnold:2022mby,Arnold:2023qwi,Arnold:2024bph,Arnold:2024whj}.

Let us first discuss the jet quenching parameter in Eq.~\eqref{eq:qhat_OE} in more detail to single out the impact of the wake.
Since medium recoils are relatively soft ($\sim T$) compared to the hard initial perturbation ($p_0\gg T$), we separate the momentum broadening of the jet from recoils by introducing a lower momentum cutoff%
\footnote{This infrared cutoff should not be confused with an often employed ultraviolet cutoff in definitions of $\hat q$ in the limit of $p_0\to\infty$, see, e.g., Ref.~\cite{Boguslavski:2023waw}.}
$\Lambda_{\min}$ using\footnote{Analogously, we can definite other transport coefficients \cref{eq:LB_qhat_L,eq:LB_qhat_parallel,eq:LB_drag,eq:LB_e_loss,eq:LB_Gamma} with a cutoff $\Lambda_\text{min}$.}
\begin{equation}\label{eq:qhat_cutoff}
    \hat q_\perp(t,\Lambda_{\min})=\partial_t \langle(\Delta\p_\perp)^2\Theta(p-\Lambda_{\min})\rangle^{(1)}\,.
\end{equation}
\Cref{eq:qhat_OE} then becomes
\begin{equation}
\label{eq:qhat_OE_cut}
    \begin{split}
        \hat q_\perp(\Lambda_{\min}) & \approx \frac{1}{2p_0}\sum_{bcd}\int_{k,p'>k'}(2\pi)^4\delta^4(p_0+k-p'-k')\\
        &\times|\mathcal M^{i_0b}_{cd}(\p_0,\k;\p',\k')|^2 \,n_b(k)(1\pm n_d(k')) \\
        & \times [{\p'_\perp}^2\Theta(p'-\Lambda_{\min})+{\k'_\perp}^2\Theta(k'-\Lambda_{\min})\\
        &\quad-\k_\perp^2\Theta(k-\Lambda_{\min})]\,.
    \end{split}
\end{equation}
In the limit $p_0>\Lambda_{\min}\gg T$, the term with $k>\Lambda_{\min}$ is exponentially suppressed as $n_b(k\gg T)\to0$. Similarly, the term that involves $k'>\Lambda_{\min}$ implies $n_d(k'\gg T)\to0$, and thus Eq.~\eqref{eq:qhat_OE_cut} simplifies to
\begin{align}
\label{eq:qhat_OE_cut_limit}
        \hat q_\perp(\Lambda_{\min}) & \approx \frac{1}{2p_0}\sum_{bcd}\int_{k,p'>k'}(2\pi)^4\delta^4(p_0+k-p'-k') \nonumber\\
        &\times|\mathcal M^{i_0b}_{cd}(\p_0,\k;\p'\k')|^2\, n_b(k) \nonumber\\
        & \times \left[ (1\pm n_d(k')){\p'_T}^2\Theta(p'-\Lambda_{\min})\right. \nonumber\\
        &\quad+\left. {\k'_T}^2\Theta(k'-\Lambda_{\min})\right]\,.
\end{align}
Even though we introduced an explicit cutoff, the additional term including $\k_T'$ remains, which shows the potential importance of the recoiling particles for the broadening. 
However, this term turns out to be subdominant. For instance, one can see this for the cutoff $\Lambda_{\min}=p_0/2$, where one always has $\Theta(p'-\Lambda_{\min})=1$, and thus the only difference to the usual expression \eqref{eq:qhat_stand} is the term ${\k'_T}^2\Theta(k'-\Lambda_{\min})$, which only contributes when $k'>p_0/2$. Due to the requirement $p'>k'$, this also implies that $p'>p_0/2$ is large. 
If $p'\approx p$, then by momentum conservation  $k$ is also large and this region is exponentially suppressed by the Bose-Einstein distribution $n_b(k)$. If $k'\approx p'\approx p_0/2$, this corresponds to large momentum exchange with a small matrix element. So indeed for $\Lambda_{\min}=p_0/2$ the standard expression \cref{eq:qhat_stand} is recovered.
However, we emphasize that
for a general cutoff $\Lambda_{\min}$, this is not guaranteed and the more general form \eqref{eq:qhat_OE_cut} should be used. Therefore, we will compare in the following our analytical and numerical results for different cutoffs $\Lambda_{\min}$.

\subsection{Comparison to kinetic theory simulations}\label{sec:resultstransport}

Having discussed the analytical expressions for the transport coefficients, we now present our numerical results 
computed using kinetic theory simulations of a minijet evolution and a direct Monte-Carlo evaluation of the corresponding single-hit formulas \cref{eq:LB_qhat,eq:LB_qhat_L,eq:LB_qhat_parallel,eq:LB_drag,eq:LB_e_loss,eq:LB_Gamma}. As discussed in the previous section, the traditional definitions of transport coefficients neglect the low momentum medium recoil, while EKT evolves partons at all momentum scales. Therefore, it is natural to introduce a low momentum cutoff $\Lambda_\text{min}$ as in \cref{eq:qhat_cutoff} to distinguish the minijet contribution from the plasma and to study how the transport coefficients depend on it.

In kinetic theory simulations, we therefore initialize a gluon minijet with energy $E=50 T$ as a narrow Gaussian perturbation with the width $\sigma=0.005E$ centered around $\p_0=(0,0,E)$
\begin{equation}
\label{eq:perturbation}
    \delta f_g(\p)=\epsilon \frac{(2\pi)^\frac{3}{2}}{(\sigma/T)^3}\frac{p_0}{p}e^{-\frac{(\bm p-\bm p_0)^2}{2\sigma^2}}\,.
\end{equation}
For $\sigma \to 0$, the perturbation \cref{eq:perturbation} corresponds to the $\delta$-distribution in the corresponding initial condition \cref{eq:init_delta} for our semi-analytical calculations. 
The initial number and energy densities are indeed given by $\delta n_0=\epsilon \nu_{g} T^3 {\rm Erf}(E/(\sqrt2\sigma))\approx \epsilon \nu_g T^3$, and $\delta e_0= \delta T^{00}_0\approx \epsilon\nu_{g}E T^3$.

\begin{figure}
    \centering
    \includegraphics{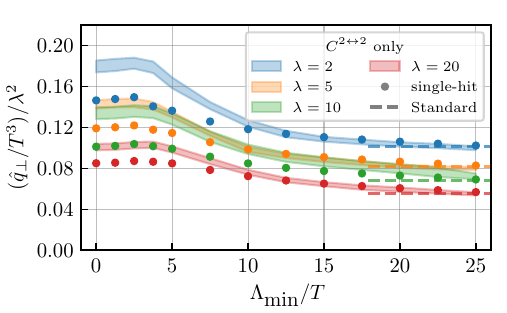}
    \caption{Transverse momentum broadening coefficient $\hat{q}_{\perp}$ for initial minijet energy $E=50T$ and different couplings $\lambda$ as a function of lower momentum cutoff $p>\Lambda_\text{min}$. Bands indicate statistical uncertainty of a single-step evolution of a gaussian perturbation in EKT, while points show the direct Monte-Carlo evaluation of a single-hit transport integrals, see \cref{eq:LB_qhat,eq:qhat_cutoff}. Dashed lines show the traditional cutoff independent definition of $\hat q_\perp$, \cref{eq:qhat_stand}.}
    \label{fig:qhat_cutoff_lambda}
\end{figure}

In \cref{fig:qhat_cutoff_lambda}, we show the results for the transverse momentum broadening coefficient $\hat{q}_{\perp}(\Lambda_{\textrm{min}})$ for different coupling constants $\lambda=2,5,10,20$ and IR cutoffs $0\leq \Lambda_\text{min}\leq 25T=E/2$. We scale out the leading coupling dependence and plot the ratio $\hat q_\perp/\lambda^2$. The bands correspond to the statistical uncertainty of ten EKT simulations, where we evaluate the EKT collision kernel for a single time step and compute the transverse momentum broadening with a given cutoff.%
\footnote{The subsequent time steps in EKT will lead to time-dependent effective transport coefficients, see \cref{sec:resultsthermalization}.} 
The small value of $\sigma=0.005E$ requires a fine discretization of the momentum grid, which is discussed in \cref{app:convergence}. We have tested the convergence of our results for different values of $\sigma$ and various grid discretizations.

The points in \cref{fig:qhat_cutoff_lambda} correspond to the single-hit evaluation of transport integrals, \cref{eq:qhat_cutoff}, for which the statistical uncertainty is negligible. For delta-like minijets, the collinear splittings do not contribute to transverse momentum broadening. For the gaussian approximation of the minijet used in kinetic simulations, the collinear splitting contribution reduces with decreasing width $\sigma$ and vanishes in the limit $\sigma\rightarrow 0$ (not shown). For a more direct comparison between the two methods, we explicitly turned off the inelastic processes to obtain \cref{fig:qhat_cutoff_lambda}.

First, one observes in \cref{fig:qhat_cutoff_lambda} that $\hat{q}_{\perp}(\Lambda_{\textrm{min}})$ scales only approximately with $\lambda^2$, as the coupling enters the screening of the matrix elements, such that larger screening effects lead to smaller values of $\hat q_{\perp}/\lambda^2$. We see a clear two-stage dependence on the IR cutoff $\Lambda_\text{min}$. For small cutoffs, the broadening of the medium recoil partons leads to an increase of the $\hat q_{\perp}$ values. For larger IR cutoffs,
momentum broadening slowly decreases and approaches the standard values of $\hat q_{\perp}$ in \cref{eq:qhat_stand} for the cutoffs $\Lambda_\text{min}\sim E/2$. The kinetic and single-hit values produce the same qualitative dependence on the cutoff and agree well for the momentum cutoffs $\Lambda_\text{min} \gtrsim 10T$.  For smaller cutoffs we observe systematically $\sim 20 \%$ larger values for EKT simulations than from the direct evaluations. In principle, for narrow gaussian perturbations $\sigma\to 0$, the two approaches should be 
equivalent. However, the EKT simulations require sufficient momentum discretization to resolve the minijet perturbation, and it becomes numerically challenging to take this limit. As discussed in \cref{app:convergence}, we have performed a systematic variation of the simulation parameters, but were not able to make the two computations completely overlap. We note that other transport coefficients seem to be less sensitive and show better agreement with direct computations even at low $\Lambda_\text{min}$. 

\begin{figure}
    \centering
    \includegraphics{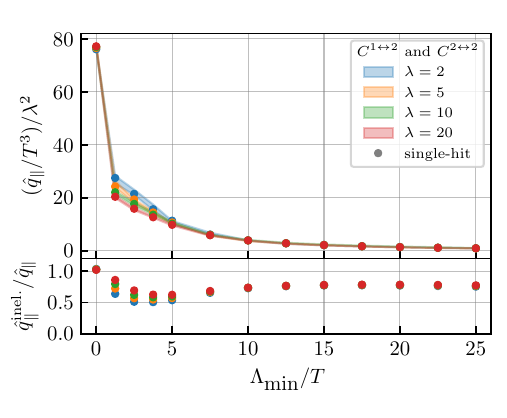}
    \caption{Longitudinal momentum broadening coefficient $\hat{q}_\parallel$ for initial minijet energy $E=50T$ and different couplings $\lambda$ as a function of lower momentum cutoff $p>\Lambda_\text{min}$. Bands indicate statistical uncertainty of a single-step evolution of a gaussian perturbation in EKT, while points show the direct Monte-Carlo evaluation of a single-hit transport integrals, see \cref{eq:LB_qhat_parallel}. The lower panel shows the relative contribution of inelastic processes.
    }
\label{fig:qhat_L_sub_cutoff_lambda}
\end{figure}

\begin{figure}
    \centering
    \includegraphics{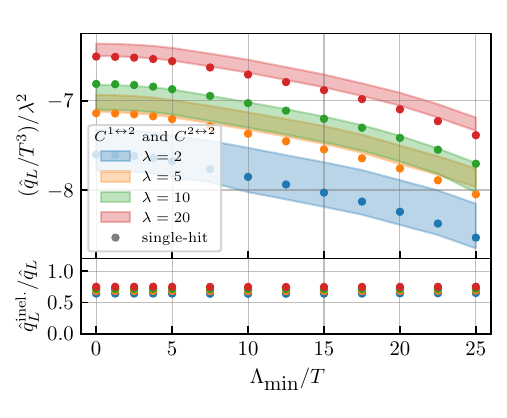}
    \caption{Longitudinal momentum broadening coefficient $\hat{q}_L$ for initial minijet energy $E=50T$ and different couplings $\lambda$ as a function of lower momentum cutoff $p>\Lambda_\text{min}$. Bands indicate statistical uncertainty of a single-step evolution of a gaussian perturbation in EKT, while points show the direct Monte-Carlo evaluation of a single-hit transport integrals, see \cref{eq:LB_qhat_L}. The lower panel shows the relative contribution of inelastic processes.
    }
\label{fig:qhat_L_cutoff_lambda}
\end{figure}

Next, we turn to longitudinal broadening. Since the initial perturbation carries a large longitudinal momentum, two types of transport coefficients can be distinguished: broadening relative to the initial jet momentum and the absolute longitudinal broadening, characterized by $\hat q_{\parallel}$ and $\hat q_{L}$, respectively, as also distinguished in \cref{eq:p_obs}.%
\footnote{As the jet thermalizes, relative longitudinal broadening becomes ambiguous, and in \cref{sec:resultsthermalization} we will consider absolute longitudinal broadening instead.}

In \cref{fig:qhat_L_sub_cutoff_lambda}, we show the relative longitudinal broadening coefficient $\hat{q}_{\parallel}(\Lambda_{\textrm{min}})$ as a function of the IR-cutoff $\Lambda_{\textrm{min}}$ for different couplings.
The collinear processes significantly modify the longitudinal momentum and are therefore included in this computation. In the lower panel, we show the relative contribution of inelastic processes to $\hat{q}_\parallel$. 
For small cutoffs $\Lambda_{\textrm{min}} \lesssim 2.5T$, the soft collinear radiation fully dominates and we observe large positive values of $\hat{q}_{\parallel}$, which reduce drastically for larger cutoffs. We emphasize that even at these larger cutoffs, the inelastic processes contribute more than half of the total relative longitudinal broadening.
We note that $\hat{q}_{\parallel}$ scales well with $\lambda^2$ and shows good agreement between EKT and single-hit computations.

Finally, in \cref{fig:qhat_L_cutoff_lambda}, we show the absolute longitudinal broadening coefficient $\hat{q}_{L}(\Lambda_{\textrm{min}})$ as a function of the IR-cutoff $\Lambda_{\textrm{min}}$. Note that here we consider the change of the absolute width $\langle\p^2_\parallel\rangle$, which is initially finite and large. Since energy is transported to lower momenta, $\langle\p^2_\parallel\rangle$ decreases and $\hat q_{L}$ is negative. 
Furthermore, for larger $\Lambda_{\textrm{min}}$, we integrate over a smaller phase space around the initial perturbation and more energy flows to lower momenta $p<\Lambda_{\textrm{min}}$. Therefore, we observe an increase of $|\hat{q}_{L}|$ with $\hat{q}_{L}<0$ 
for larger $\Lambda_{\textrm{min}}$.
We also find a similar ordering of the absolute values 
$|\hat{q}_{L}|/\lambda^2$ that grow for smaller couplings as for $\hat q_\perp$ in \cref{fig:qhat_cutoff_lambda}. We note that independently of the cutoff, the inelastic processes contribute more than half to this transport coefficient, as shown in the lower panel.
Overall, we see a good agreement between kinetic and single-hit computations within statistical uncertainties.

\begin{figure}
    \includegraphics{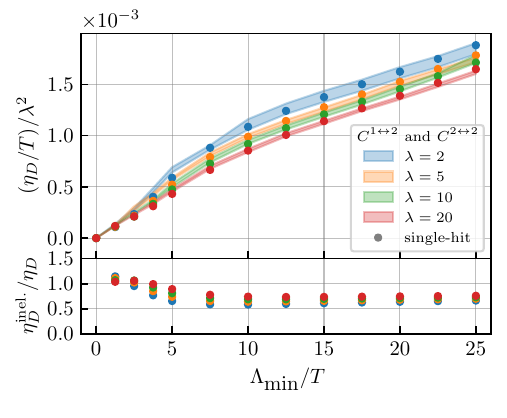}
    \caption{Longitudinal momentum drag coefficient $\eta_D$    for initial minijet energy $E=50T$ and different couplings $\lambda$ as a function of lower momentum cutoff $p>\Lambda_\text{min}$. Bands indicate statistical uncertainty of a single-step evolution of a gaussian perturbation in EKT, while points show the direct Monte-Carlo evaluation of a single-hit transport integrals, see \cref{eq:LB_drag}. The lower panel shows the relative contribution of inelastic processes.}
    \label{fig:drag_cutoff_lambda}
\end{figure}

Finally, in \cref{fig:drag_cutoff_lambda} we show results for the drag coefficient $\eta_D(\Lambda_{\textrm{min}})$ given by \cref{eq:LB_drag}, which measures the longitudinal momentum loss. 
Due to momentum conservation, the net longitudinal momentum is conserved yielding $\eta_D(\Lambda_{\textrm{min}}{=}0)=0$. For larger cutoffs, more momentum can flow to $p<\Lambda_{\textrm{min}}$, and the drag increases. From the lower panel of \cref{fig:drag_cutoff_lambda} we see that the inelastic processes are the dominant source of the drag. For small $\Lambda_{\textrm{min}}$, the ratio $\eta_D^{\mathrm{inel.}}/\eta_D$ is slightly larger than one. Therefore, at those small cutoffs, the elastic processes contribute negatively to the (very small) drag coefficient. For this transport coefficient, we see a good agreement between single-hit and EKT computations. We also computed the energy loss coefficient $\hat e$. However, for a narrow jet, the longitudinal momentum and energy loss are related by a jet energy factor and a minus sign (results not shown). 

In summary, we find significant cutoff dependence of all transport coefficients, indicating a non-negligible medium recoil contributions.
For $\hat{q}_{\perp}$ only elastic scatterings contribute, but other transport coefficients (see \cref{fig:qhat_L_sub_cutoff_lambda,fig:qhat_L_cutoff_lambda,fig:drag_cutoff_lambda}) are dominated by radiative processes (ratios in the lower panels greater then $0.5$). Therefore the complete description of energy and longitudinal momentum transport must necessarily include inelastic processes.

\section{Minijet thermalization time}
\label{sec:resultsthermalization}

\subsection{Early time behavior and $\hat q$ scaling }

In this section, we study the relation between jet transport coefficients and the thermalization time of the leading jet parton. Following~\cite{Mehtar-Tani:2022zwf,Zhou:2024ysb}, we initialize the Yang-Mills kinetic theory with a Bose-Einstein distribution $n_\text{eq}$ with temperature $T$ for the background medium and add a narrow Gaussian perturbation with the energy $|\p_0|=E$ given by \cref{eq:perturbation} that initializes the linearized Boltzmann equation \eqref{eq:LB} for the jet evolution. In this section, we use initial perturbations of width $\sigma = 0.01E$ (see \cref{app:convergence} for further discretization details).
At late times, the perturbation equilibrates to a boosted thermal distribution
\begin{equation}\label{eq:deltafthermal}
    \delta f_g^\text{\tiny (eq)}(t,\bm p)=\left.(\delta T\partial_T+\delta u^z\partial_{u^z})n_\text{eq}(p_\mu u^\mu)\right|_{u^z=0}\,.
\end{equation}
Correspondingly, the equilibrium number and energy densities are $\delta n_\text{eq} = \nu_g \frac{3\zeta(3)}{\pi^2}T^2 \delta T$ and $\delta e_\text{eq} = \nu_g \frac{4\pi^2}{30}T^3 \delta T$.
Energy and momentum conservation ensures a relation between initial and final energy densities, and we find the net-temperature increase%
\footnote{Note that the velocity term in \cref{eq:deltafthermal} does not contribute to the energy density integral (or any integral even in $p_z$).}
\begin{equation}
   \delta T= \frac{30}{4\pi^2}\,\epsilon E,
\end{equation}
which for $\epsilon \ll 1$ can be always made arbitrarily small, i.e., $\delta T\ll T$. From this, we can also compute the transverse and longitudinal momentum expectation values for equilibrated partons according to \cref{eq:exp_value}
\begin{equation}\label{eq:pL2_pT2_eq}
      \langle \p_{\parallel}^2\rangle_{\rm eq}=\tfrac12\langle \p_{\perp}^2\rangle_{\rm eq} = \nu_g\frac{20}{\pi^2}\zeta(5)T^4\delta T/\delta n_0\,,
\end{equation}
with the initial jet parton density $\delta n_{0} = \nu_g \epsilon T^3$.

\begin{figure}
    \centering
    \includegraphics{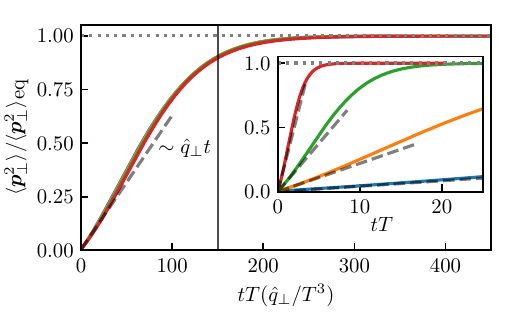}
    \includegraphics{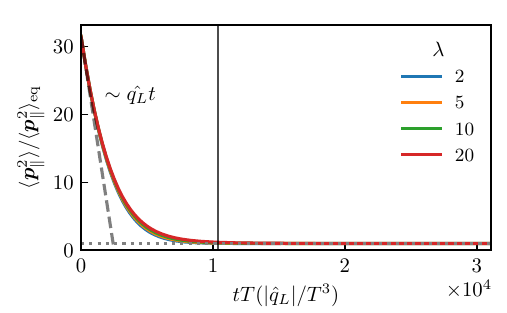}
    \caption{Momentum broadening $\langle \p_{\perp}^2 \rangle$ and $\langle \p_{\parallel}^2 \rangle$ as a function of time for different couplings $\lambda$, at fixed jet energy $E=50T$. We scale the time with transverse and longitudinal broadening coefficients $\hat{q}_{\perp}$ and $\hat{q}_{L}$, respectively. The gray lines show the linear slope. The inset shows the unscaled evolution as a function of $tT$. The vertical black line indicates the time where the minijet is within $10\%$ deviation from the equilibrium value.}
    \label{fig:pT2_L}
\end{figure}

In order to investigate the equilibration of the initial jet perturbation, we will study its broadening in momentum space. In the upper panel of \cref{fig:pT2_L}, we show the squared transverse momentum evolution relative to the equilibrium values for different couplings $\lambda$. 
The jet perturbation has an initially small transverse width, $\langle \p_{\perp}^2\rangle_0\approx 0$. During the evolution, the perturbation isotropizes.
In the inset we display the evolution of $\langle \p_{\perp}^2\rangle$ as a function of time in units of (constant) background temperature $T$, i.e., $tT$. For small $\lambda$, the system takes longer to reach equilibrium. The initial increase of $\langle \p_{\perp}^2\rangle$ is well captured by the straight line $\langle \p_{\perp}^2\rangle\approx \hat{q}_\perp t$ (dashed lines) following the single-hit expansion. Therefore, in the main panel of the figure, we rescale the time with $\hat{q}_\perp$ computed from \cref{eq:LB_qhat} and shown as points in \cref{fig:qhat_cutoff_lambda} for $\Lambda_{\min} = 0$. 
At intermediate times, inelastic processes will also contribute and the transverse width increases at a rate faster than the initial linear growth. 
We note that after rescaling, all of the curves collapse onto each other at early times and also, remarkably, at late times. 
This indicates that in our settings, the coupling dependence of the minijet equilibration at all times is reasonably well captured by the transport coefficient $\hat{q}_{\perp}$. We will discuss late-time deviations not visible in \cref{fig:pT2_L} in the next section. Compared to a previous study \cite{Zhou:2024ysb}, where the late time transport coefficient $\eta/s$ was used to rescale time, the ``early time'' quantity $\hat{q}_\perp$ provides a better description.

In the lower panel of \cref{fig:pT2_L}, we show the evolution of the longitudinal momentum. The jet perturbation has an initially large longitudinal momentum $\langle \p_{\parallel}^2\rangle_0\approx E^2$, that rapidly decays to a thermal value. After rescaling the time axis with the single-hit $|\hat q_\parallel|$, we see a reasonably good collapse for different couplings even for late times.
However, we note that in contrast to transverse broadening, the decrease in the squared longitudinal momentum is dominated by the inelastic processes. Although on the scale of \cref{fig:pT2_L}, $\hat q_\parallel$ describes the initial slope of the curve reasonably well (dashed line), the curve actually deviates from this linear behavior earlier than for transverse momentum broadening. We have observed this by considering 
the time dependence of the derivative of the curve (data not shown).

\begin{figure}
    \centering
    \includegraphics{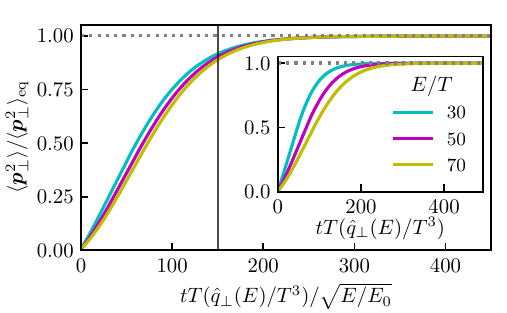}
    \caption{Transverse momentum broadening $\langle \p_{\perp}^2 \rangle$ for coupling $\lambda=10$ and different initial minijet energy $E$. We scale the time axis by $\hat q_{\perp}(E)$ and $\sqrt{E}$. The vertical black line indicates the time where the minijet is within $10\%$ deviation from the equilibrium value. The inset shows evolution where time is only scaled by $\hat{q}_{\perp}(E)$.}
    \label{fig:pT2_E}
\end{figure}

Finally, in \cref{fig:pT2_E}, we show the time evolution of the transverse momentum squared $\langle \p_{\perp}^2\rangle$ for different jet energies $E$ for $\lambda=10$. The inset shows that higher-energy jets take longer to thermalize. The parametric scaling of the thermalization time with energy can be obtained by considering the time it takes the initial jet to radiate its energy away~\cite{Kurkela:2014tea}. The soft splitting rate is given by $\frac{\rmd N}{\rmd z\rmd t}\approx\frac{\alpha_s}{2\pi}\frac{2C_A}{z}\sqrt{\frac{C_A\hat{q}_{\perp}}{zE}}$ where $z$ is the radiated momentum fraction. Then the expected energy loss $\Delta E$ from a single radiation over time $\Delta t$ is $\Delta E=\Delta t \int\rmd z zE\frac{\rmd N}{\rmd z\rmd t}$. Requiring $\Delta E \sim E$ results in the thermalization time scaling $t_\text{th}\sim\frac{1}{\alpha_s}\sqrt{E/\hat{q}_{\perp}}$. Using that parametrically $\hat{q}_{\perp} \sim \alpha_s^2 T^3$~\cite{Ghiglieri:2015zma},  we reproduce the parametric behavior of the thermalization time $t_\text{th}\sim \frac{1}{\alpha_s^2 T} \sqrt{E/T}$, see Ref.~\cite{Kurkela:2014tea}. On the other hand, we can express the thermalization time without explicit dependence on the coupling constant as 
\begin{align}
    \label{eq:tthE}
    t_\text{th}^E\sim \frac{1}{T} \frac{\sqrt{E/E_0}}{\hat{q}_{\perp}/T^3}\,,
\end{align}
where $E_0$ is some fixed reference energy scale. In the main panel of \cref{fig:pT2_E} we rescale the time using \cref{eq:tthE}
with $E_0=50 T$. 
The initial slope of the curve no longer agrees as in \cref{fig:pT2_L}, but all curves corresponding to different energies approach equilibrium simultaneously.

Knowing how the evolution scales with the parameters $\hat{q}_{\perp}$ and $E$, we can now give a quantitative estimation of the equilibration time of the minijet. We choose the thermalization time $t_{\textrm{th}}$ to be the one where $\langle \p_{\perp}^2\rangle$ reaches $90\%$ of the equilibrium value (vertical black line in \cref{fig:pT2_E})
\begin{align}\label{eq:eq_condition}
    \langle \p_{\perp}^2\rangle(t = t_{\textrm{th}}) &\equiv 0.9\langle \p_{\perp}^2\rangle_{\textrm{eq}}.
\end{align}
In our kinetic theory simulations, we find that a minijet with initial energy $E=50T$ thermalizes at the time
\begin{equation}
    t_{\mathrm{th}}\approx 150\, T^{-1}\left( \frac{\hat{q}_{\perp}}{T^3}\right)^{-1}\left( \frac{E}{E_0}\right)^{1/2},
\end{equation}
To provide a phenomenological estimate of the thermalization time in heavy-ion collisions, we choose the constant medium temperature $T=0.3\,\text{GeV}$ and $E=15\,\text{GeV}$,
\begin{equation}
    t_{\mathrm{th}}\approx \frac{100\,\textrm{fm}}{\hat{q}_{\perp}/T^3}\left( \frac{T}{0.3\,\mathrm{GeV}}\right)^{-1}\left( \frac{E}{15\,\mathrm{GeV}}\right)^{1/2}.\label{eq:th}
\end{equation}
For typical values of the jet quenching coefficient $\hat q_{\perp}/T^3= 2-10$ ($\hat q_{\perp} = 0.3-1.4\,\text{GeV}^2/\text{fm}$)~\cite{JETSCAPE:2024cqe}, we get $t_\text{th}=50 - 10\,\text{fm}$.
This is a significantly longer time than previous estimates based on extrapolations using shear viscosity over entropy ratio $\eta/s$~\cite{Zhou:2024ysb}. 
Note that in leading order QCD kinetic theory, the coupling constant does not run, and the low and high momentum partons interact with the same coupling strength. The phenomenologically favored values of $\eta/s\sim 0.16$ correspond to a more strongly coupled plasma ($\lambda>20$) than the phenomenologically favored values of $\hat q_{\perp}/T^3\sim 2-10$ ($\lambda\sim 5-10$). This reflects the fact that higher-momentum partons interact more weakly than the bulk QGP. Therefore, the jet thermalization time obtained in \cref{eq:th} is more realistic and consistent with jet phenomenology.

\subsection{Late time relaxation and second-order hydrodynamics}
Strictly speaking,  \cref{eq:th} provides only a time estimate of a minijet to lose most of the initial state information. The final approach to equilibrium and thermalization are governed by the universal hydrodynamic relaxation applicable to any small perturbation around equilibrium.
Here, we will study the late-time approach of minijet perturbations to equilibrium, \cref{eq:deltafthermal}. In equilibrium, the energy-momentum tensor is given by
\begin{equation}\label{eq:obs_eq}
    \begin{split}
    \delta T^{00}_{\rm eq} &= \nu_g \tfrac{4\pi^2}{30}T^3 \delta T\,,\\
        \delta T^{0i}_{\rm eq} &= \tfrac{2}{15}\nu_g\tfrac{\pi^2}{3}T^4\delta v^i\,,\\
        \delta T^{ii}_{\rm eq} &= \tfrac{1}{3}\delta T^{00}_{\rm eq}\,,\quad \delta T^{ij}_{\rm eq}=0\,.
    \end{split}
\end{equation}
\begin{figure}
    \centering
    \includegraphics[width=\linewidth]{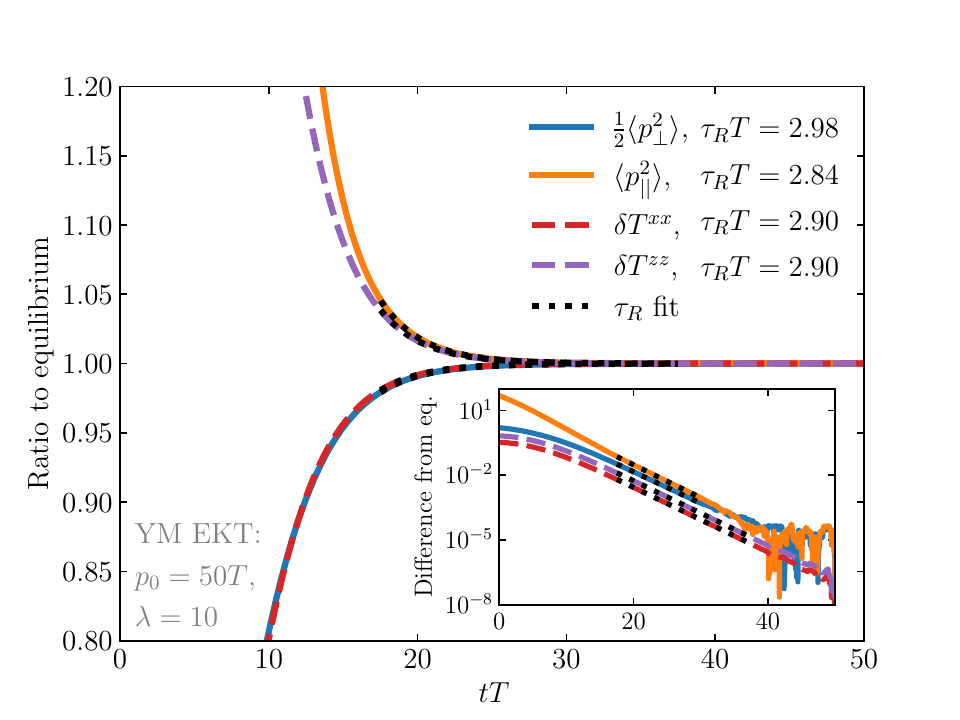}
    \caption{The late-time evolution of longitudinal and transverse pressure and momentum squared, their relaxation to equilibrium, and the extracted relaxation times.}
    \label{fig:tauR}
\end{figure}

\Cref{fig:tauR} shows the late-time evolution of longitudinal and transverse pressure and momentum, all of which asymptotically approach their equilibrium values \cref{eq:pL2_pT2_eq,eq:obs_eq}. In the inset, we show the deviations from equilibrium, revealing an exponential relaxation behavior $\propto e^{-t/\tau_R}$.
This indicates that the late time equilibration of perturbations could be well described by the relaxation time approximation of kinetic theory~\cite{Anderson:1974nyl,Rocha:2021zcw}, which is
characterized by a single relaxation time $\tau_R$.
In the legend, we show the extracted relaxation time $\tau_R$ for each observable in units of the background temperature. The fitted values of around $\tau_R T \approx 2.9$ in \cref{fig:tauR} are similar for both pressure and momentum observables.
We note that the relaxation time for longitudinal and transverse pressure are identical thanks to energy conservation and conformality of massless kinetic theory. This does not apply to averaged transverse and longitudinal momentum and we observe 5\% variation in the extracted value. 
These differences might reflect the residual momentum dependence on relaxation time in QCD kinetic theory.

The late time equilibration of the energy-momentum tensor is described by viscous hydrodynamics.  
In second-order hydrodynamics, the relaxation time of spatial components of the energy-momentum tensor is governed by
the Müller-Israel-Stewart (MIS) equation~\cite{Muller:1967zza,Israel:1979wp}
\begin{equation}
  \tau_\pi  \partial_t \pi^{ij} = -\pi^{ij}+\pi_1^{ij}\,.
\end{equation}
Here $\pi^{ij}$ is the non-ideal dissipative part of the energy-momentum tensor, $T^{ij}=\delta^{ij}\mathcal P+\pi^{ij}$, and $\pi^{ij}_1$ is the first-order gradient correction to the energy-momentum tensor.
The second-order hydrodynamic transport coefficient $\tau_\pi$ describes the relaxation of $\pi^{ij}$ to the Navier-Stokes expectation  $\pi^{ij}_1$, which is zero for our homogeneous system. $\tau_\pi$ was computed in QCD kinetic theory at leading~\cite{York:2008rr} and next-to-leading order~\cite{Ghiglieri:2018dgf}.
We can compare our extracted relaxation time $\tau_R$ to the computed values of $\tau_\pi$ in 0-flavor QCD for $\lambda=10$, where $\tau_\pi T/(\eta/s) \approx 5.2$ at $m_D/T = \sqrt{\lambda/3}\approx 1.8$ in Fig.~1 in~\cite{York:2008rr}.
For $\lambda=10$ the specific shear viscosity in a gluonic plasma is $\eta/s\approx 0.513$~\cite{Boguslavski:2024kbd}, such that the extracted value of the relaxation time corresponds to $\tau_R T/(\eta/s) \approx 5.65$. This value is in reasonable agreement (within 5\%) with the value found in the literature. Unlike the thermalization time, \cref{eq:th}, the $\tau_\pi$ does not depend on the initial minijet energy, since such information has been lost by the time the hydrodynamic equations become applicable.

\section{Conclusion}\label{sec:conclusions}

As a high-momentum parton (minjet) traverses a high-temperature QCD plasma, it interacts with the medium through scatterings characterized by jet transport coefficients, such as the transverse momentum broadening $\hat{q}_{\perp}$ and drag coefficient $\eta_D$. 
In this work, we investigated the thermalization of on-shell minijets in a weakly coupled 
gluonic plasma using AMY kinetic simulations with elastic scatterings and collinear radiation on a thermal background. In particular, our implementation employs an isotropic hard-thermal-loop (HTL) screening of soft exchanges. Because the kinetic evolution incorporates both elastic and inelastic processes as well as the recoil of medium constituents from high-momentum partons, we observe quantitative deviations between the full EKT evolution and the standard definitions of transport coefficients.

We performed a single-scattering (opacity) expansion of the Boltzmann equation and derived general expressions for the expectation values of moments of the distribution function, such as the squared transverse momentum $\langle \p_\perp^2\rangle$. By defining transport coefficients through the time derivatives of these moments, we recover the conventional expressions known from analytic treatments. However, standard formulations of jet transport coefficients typically neglect the contributions from recoiling medium particles and inelastic processes. Our analysis demonstrates that these effects are quantitatively important and must be included for a consistent description of the minijet evolution. The transport coefficients extracted from the single-hit expansion show good agreement with the early-time behavior observed in full kinetic simulations. Finally, we show that the traditional expressions can be approximately recovered by introducing a large infrared cutoff that suppresses the recoil contributions.

We also studied the thermalization of a minijet in a non-expanding thermal medium. The early-time evolution is well described by jet-transport coefficients, as expected. Remarkably, when the evolution time is rescaled by these transport coefficients, most notably by the jet quenching coefficient $\hat q_\perp$, the results exhibit a universal scaling behavior across different coupling strengths and energies, extending smoothly up to the thermalization stage and outperforming previous rescalings based on $\eta/s$. From this analysis, we derived a compact phenomenological formula that provides realistic estimates of the minijet thermalization time in heavy-ion conditions. 
Close to equilibrium, we confirm that the relaxation of perturbations is consistent with second-order hydrodynamic coefficients obtained in earlier studies.

A natural next step is to include dynamical quarks and quark–gluon conversion channels within the same kinetic framework. In \cref{app:HTL_Screening} we have already derived a practical implementation of isotropic HTL screening prescription for fermions  and validated it against analytical result of gluon-quark conversion rate in \cref{fig:conversionrate}. 
However, our preliminary attempts to extract transport coefficients from the linearized EKT evolution including fermions suffer from numerical instabilities, which currently prevent a quantitative comparison with the direct computation of transport coefficients, such as done for the gluonic plasma in \cref{sec:resultstransport}. These instabilities originate from the current implementation of the Monte Carlo sampler, which does not account for the fact that minijets have very small support in the sampled phase space. As a result, the sampling becomes highly inefficient, which is not the case for the smoother distributions encountered near equilibrium.
A revised sampling strategy tailored to such extreme initial conditions could improve numerical stability, which we leave for future work.

Another extension of this work is to relax the assumption of a static medium. Since the QGP expands rapidly, the evolving background can significantly modify the dynamics of minijet propagation. As discussed in \cite{Zhou:2024ysb} and noted in this work, such dynamical effects introduce additional linear contributions to the collision kernels that stem from the perturbation of the (screened) matrix elements. 
These terms also affect the transport properties of the minijet. However, their size and form can depend strongly on the chosen screening prescription for elastic scatterings and collinear splittings.
To our knowledge, these additional terms have not been included in previous studies involving non-equilibrium backgrounds~\cite{Boguslavski:2024kbd},
but are needed for a consistent linearized description of transport in an expanding and dynamically evolving medium.

Finally, for an evolving anisotropic medium there are further corrections that should be taken into account.
In particular, 
currently the internal soft propagators are screened isotropically.
Large momentum-anisotropy of the non-equilibrium medium affects the collision kernels and the underlying collinear splitting rate \cite{Altenburger:2025iqa, Lindenbauer:2025ctw}. While previous studies within QCD kinetic theory \cite{AbraaoYork:2014hbk, Kurkela:2014tea, Kurkela:2015qoa, Kurkela:2018oqw, Kurkela:2018xxd, Kurkela:2021ctp, Du:2020dvp, Du:2020zqg, Kurkela:2018vqr, Kurkela:2018wud, Boguslavski:2024kbd} neglected the effect of plasma instabilities resulting from the anisotropy \cite{Mrowczynski:1993qm, Mrowczynski:2016etf}, justified by the observation that they do not play a dominant role in the high-occupancy regime of the thermalization process \cite{Berges:2013eia, Berges:2013fga}, recent studies have put forward suggestions of how they can be accounted for in the descriptions of hard probes and kinetic theory \cite{Hauksson:2020wsm, Hauksson:2021okc}. Further studies are needed to obtain a practical way of their inclusion in kinetic theory simulations.

In summary, our work demonstrated how the simplified picture encoded in jet transport coefficients emerges from the full kinetic evolution of on-shell partons in the QGP. Despite the differences in the details, we demonstrated that $\hat q$ scaling provides a practical mapping from medium properties to jet quenching times. 
Confronting this with system-size and energy dependencies in light- and heavy-ion collisions can deliver tighter constraints on jet-transport coefficients and on the onset of thermalization in small systems.

\begin{acknowledgments}
We thank Ismail Soudi for useful discussions.
This work is supported by the DFG through the Emmy Noether Programme (project number 496831614) (AM, AT, FZ), through CRC 1225 ISOQUANT (project number 27381115) (AM, AT, FZ), and by the Austrian Science Fund (FWF) under Grant DOI 10.55776/P34455 (KB, FL) and 10.55776/J4902 (FL). FL is a recipient of a DOC Fellowship of the Austrian Academy of Sciences at TU Wien (project 27203). For the purpose of open access, the authors have applied a CC BY public copyright license to any Author Accepted Manuscript (AAM) version arising from this submission. We thank ECT* for support at the Workshop “Attractors and thermalization in nuclear collisions and cold quantum gases” during which we profited from useful discussions. We acknowledge support by the state of Baden-W\"urttemberg through bwHPC and the German Research Foundation (DFG) through grant no INST 39/963-1 FUGG (bwForCluster NEMO).
\end{acknowledgments}

\bibliography{refs_med,bib}

\onecolumngrid
\appendix

\section{Elastic and inelastic collision kernels in EKT}
\label{App:EKT}
The Boltzmann equation can be written in the form $\partial_t f_i(\bm p)=C_i[\{f\}]$. 
Then the collision kernel includes elastic and radiative processes, $C_i=C_i^{2\leftrightarrow2}+C^{1\leftrightarrow2}_i$, and is always local. The elastic collision kernel is given by \cite{Arnold:2002zm}
\begin{equation} \label{eq:c22}
    \begin{split}
        &C_a^{2\leftrightarrow 2}[\{f\}] = \frac 1{4p_1\nu_a} \sum_{bcd} \int d\Omega^{2\leftrightarrow2}
        \left|{\cal M}^{ab}_{cd}(\bm p_1,\bm p_2;\bm p_3,\bm p_4)\strut\right|^2\mathcal{F}(\bm p_1,\bm p_2;\bm p_3,\bm p_4) \,,
    \end{split}
\end{equation}
where the phase space integral is
\begin{equation}\label{eq:PhaseSpace_C22}
    \begin{split}
        \int d\Omega^{2\leftrightarrow2}&= \int \frac{d^3\bm p_2}{(2\pi)^3}\frac{1}{2E_2}\frac{d^3\bm p_3}{(2\pi)^3}\frac{1}{2E_3}\frac{d^3\bm p_4}{(2\pi)^3}\frac{1}{2E_4}(2\pi)^4 \delta^{4}(P_1+P_2-P_3-P_4)\,,
    \end{split}
\end{equation}
with the statistical factor
\begin{equation}
    \mathcal{F}(\bm p_1,\bm p_2;\bm p_3,\bm p_4)=f_c(\bm p_3)f_d(\bm p_4)(1\pm f_a(\bm p_1))(1\pm f_b(\bm p_2))-f_a(\bm p_1)f_b(\bm p_2)(1\pm f_c(\bm p_3))(1\pm f_d(\bm p_4))\,,
\end{equation}
where the upper (lower) sign is used if the distribution is for bosons (fermions), which reads in equilibrium $n_a(p)=1/(e^{p/T}\mp1)$. For the matrix element, we take the scattering matrix elements in vacuum, summed over all incoming and outgoing spin and color states, as tabulated in Ref.~\cite{Arnold:2002zm}. Medium effects should be included by using the resummed hard-thermal loop propagator for the internal soft quark or gluon propagators. As argued in Ref.~\cite{Arnold:2002zm}, this can be easily incorporated by a simple replacement in the $u$- and $t$- channel parts of the matrix element, which we will discuss in more detail in the next section.
The radiative collision kernel is given by
\begin{equation}
    \begin{split}\label{eq:c12}
        C_a^{1\leftrightarrow 2}[\{f\}] = \frac{(2\pi)^3}{2|\p|^2\nu_a}&\sum_{bc}\int_0^\infty dp'dk'\delta(p-p'-k')\gamma^{a}_{bc}(\p;p'\hat\p,k'\hat\p)\mathcal F(\p;p'\hat\p,k'\hat\p)\\
        -\frac{(2\pi)^3}{|\p|^2\nu_a}&\sum_{bc}\int_0^\infty dp'dk'\delta(p+k'-p')\gamma^c_{ab}(p'\hat\p;\p,k'\hat\p)\mathcal F(p'\hat\p;\p,k'\hat\p)\,,
    \end{split}
\end{equation}
with the statistical factor
\begin{equation}
    \mathcal F(\p_1;\p_2,\p_3) = f_b(\p_2)f_c(\p_3)(1\pm f_a(\p_1))-f_a(\p_1)(1\pm f_b(\p_2))(1\pm f_c(\p_3))\,.
\end{equation}
This radiation kernel describes strict collinear splitting (where all momenta are proportional to a unit vector $\vb {\hat n}$), and is given by the AMY rates \cite{Arnold:2002ja, Arnold:2002zm}, where the effective splitting matrix elements $\gamma$ are given by
\begin{align}
    \gamma^{q}_{qg}(p;p',k)&=\gamma^{\bar q}_{\bar q g}(p; p', k)=\frac{p'^2+p^2}{p'^2p^2k^3}\mathcal F^{\hat n}_q(p,p',k)\,,\\
    \gamma^g_{q\bar q}(p; p', k)&=\frac{k^2+p'^2}{p'^3p^3k^3}\mathcal F^{\hat n}_q(k,-p',p)\,,\\
    \gamma^{g}_{gg}(p; p',k)&=\frac{p'^4+p^4+k^4}{p'^3p^3k^3}\mathcal F^{\hat n}_g(p,p',k)\,,
\end{align}
where
\begin{align}
    \mathcal F^{\hat n}_s(p,p,k)=\frac{d_s C_s\alpha_s}{2(2\pi)^3}\int\frac{\dd[2]{\vb h}}{(2\pi)^2}2\vb h\cdot \RE \vb F_s^{\hat n}(\vb h; p',p,k)\,,
\end{align}
and $\vb F$ is the solution to the integral equation
\begin{align}\label{eq:fullsplitting}
    \begin{split}
    2\vb h&=i\delta E(\vb h; p',p,k)\vb F_s^{\hat n}(\vb h; p', p,k)+g^2\int\frac{\dd[2]{\vb q_\perp}}{(2\pi)^2}\mathcal A(\vb q_\perp)\\
    &\times\Bigg\{(C_s-\tfrac{C_A}{2})[\vb F^{\hat n}(\vb h; p',p,k)-\vb F_s^{\hat n}(\vb h-k\vb q_\perp; p',pk)]\\
    &+\tfrac{C_A}{2}[\vb F_s^{\hat n}(\vb h; p',p,k)-\vb F_s^{\hat n}(\vb h+p'\vb q_\perp; p',p,k)]\\
    &+\tfrac{C_A}{2}[\vb F_s^{\hat n}(\vb h; p',p,k)-\vb F_s^{\hat n}(\vb h-p\vb q_\perp; p',p,k)]\Bigg\}\,.
    \end{split}
\end{align}
The vector $\vb h$ is a two-dimensional vector in the transverse plane to the direction of the splitting particles, $\vb {\hat n}$. The elastic collision kernel $\mathcal A(\vb q_\perp)$ encodes the broadening of hard particles during the splitting process. It can be represented as a Wightman correlator of the gluon field generated by the hard particles moving in the plasma. With an isotropic screening approximation, it can be simplified to the simple sum rule result \cite{Aurenche:2002pd}
\begin{align}
    \mathcal A(\vb q_\perp)=T_\ast\left(\frac{1}{q_\perp^2}-\frac{1}{q_\perp^2+m_D^2}\right),
\end{align}
where the effective infrared temperature $T_\ast$ is given by
\begin{align}
    T_\ast = \frac{\sum_s\nu_s \frac{g^2C_s}{d_A}\int\frac{\dd[3]{\p}}{(2\pi)^3}f(\p)(1\pm f(\p))}{m_D^2}
\end{align}
and with the nonequilibrium Debye mass
\begin{align}
	m_D^2 =  \sum_s4\nu_s\frac{g^2C_s}{d_A}\int\frac{\dd[3]{\p}}{(2\pi)^32|\p|}f_s(\p).
\end{align}
Note that in practice, we do not solve \cref{eq:fullsplitting} directly. Instead we interpolate between the analytically known Bethe-Heitler and Landau-Pomeranchuk-Migdal limits of the collinear splitting rates~\cite{Berges:2020fwq}. This reproduces well the results of
\cref{eq:fullsplitting} as shown in Appendix B of Ref.~\cite{Kurkela:2022qhn}.

Throughout the paper, we use the Boltzmann equation linearized around thermal equilibrium, $f=n+\delta f$, (see Eq.~\eqref{eq:LB}). The linearized collision kernels are simply given by the original ones \eqref{eq:c22} and \eqref{eq:c12}, with replaced statistical factors $\mathcal F\mapsto\delta\mathcal F$,
\begin{equation}
    \begin{split}
        \delta\mathcal{F}(\bm p_1,\bm p_2;\bm p_3,\bm p_4) 
        &= \delta f_c(\bm p_3) \left[n_d(p_4)(1 \pm n_a(p_1))(1 \pm n_b(p_2)) \mp_c n_a(p_1)n_b(p_2)(1\pm n_d(p_4))\right] \\
        &+ \delta f_d(\bm p_4) \left[n_c(p_3)(1 \pm n_a(p_1))(1 \pm n_b(p_2)) \mp_d n_a(p_1)n_b(p_2)(1\pm n_c(p_3))\right] \\
        &- \delta f_a(\bm p_1) \left[n_b(p_2)(1 \pm n_c(p_3))(1 \pm n_d(p_4)) \mp_a n_c(p_3)n_d(p_4)(1\pm n_b(p_2)) \right] \\
        &- \delta f_b(\bm p_2) \left[n_a(p_1)(1 \pm n_c(p_3))(1 \pm n_d(p_4)) \mp_b n_c(p_3)n_d(p_4)(1\pm n_a(p_1)) \right] \,,
    \end{split}
\end{equation}
and
\begin{equation}
    \begin{split}
        \delta\mathcal F(\p_1;\p_2,\p_3) &= \delta f_b(\p_2)[n_c(p_3)(1\pm n_a(p_1))\mp_bn_a(p_1)(1\pm n_c(p_3))]\\
        &+ \delta f_c(\p_3)[n_b(p_2)(1\pm n_a(p_1))\mp_c n_a(p_1)(1\pm n_b(p_2))]\\
        &- \delta f_a(\p_1)[(1\pm n_b(p_2))(1\pm n_c(p_3))\mp_a n_b(p_2)n_c(p_3)]\,.
    \end{split}
\end{equation}
Additionally, the matrix element linearization can be found in~\cite{Zhou:2024ysb}. We linearize around thermal equilibrium, and thus the term with matrix-element linearization vanishes as statistical factors cancel exactly.

\section{Isotropic HTL screening for soft gluon and quark exchange}
\label{app:HTL_Screening}
Here, we describe the full HTL screening of the 2$\leftrightarrow$2 matrix element, and show that our implementation reproduces the known analytical results, while the commonly used vacuum LO pQCD matrix elements with Debye-like screening prescription are insufficient to simultaneously describe all transport coefficients. The implementation of the HTL prescription for soft-gluon exchange has been described before \cite{Boguslavski:2024kbd}, which we reiterate here for completeness and extend it for soft-fermion exchange. 

According to the prescription outlined in Ref.~\cite{Arnold:2002zm}, for gluons, one needs to replace
\begin{align}
    \label{eq:amy_replacement}
    {\frac{(s-u)^2}{\underline{t^2}}} &\to  \left|G^R_{\mu\nu}(P-P')\;(P+P')^\mu(K+K')^\nu\right|^2\,.
\end{align}
While this prescription is gauge-invariant \cite{Boguslavski:2024kbd}, we use here the form of the retarded gluon propagator in strict Coulomb gauge~\cite{Bellac:2011kqa, Ghiglieri:2020dpq},
\begin{subequations}
    \begin{align}
        G^{00}(Q) &= \frac{1}{q^2+\Pi^{00}(\omega/q)}\,, &
        G^T(Q) &= \frac{-1}{q^2-\omega^2+\Pi^T(\omega/q)}\,,
    \end{align}
\end{subequations}
with $G^{ij}(Q)=\left(\delta^{ij}-\frac{q^iq^j}{q^2}\right)G^T(Q)$, and the self-energies are given by
\begin{subequations}
    \begin{align}
        \RE{\Pi}^{00}(x) &= m_D^2\left(1-\frac{x}{2}\ln\left|\frac{x+1}{x-1}\right|\right)\,, & 
        \IM\Pi^{00}(x) &= \frac{xm_D^2\pi}{2}\theta(1-|x|)\,,\\
        \RE \Pi^T(x) &= \frac{m_D^2}{2}-\frac{1}{2}(1-x^2)\RE\Pi^{00}(x)\,,&
        \IM\Pi^T(x) &= -\frac{1}{2}(1-x^2)\IM\Pi^{00}(x)\,,
    \end{align}
\end{subequations}
where $x=\omega/q$. Kinematically, $|x|=|\omega|/q < 1$, and thus the imaginary part is always nonzero for $\omega\neq 0$. Here, we define $Q=P'-P=(\omega,\q)$. Furthermore, $G(-Q)$ corresponds to the advanced propagator, and $\IM\Pi(-Q)=-\IM\Pi(Q)$. From this, it is straightforward to obtain~\cite{Boguslavski:2024kbd, Boguslavski:2023waw}
\begin{align}
    \label{eq:HTL_propagator_explicit_expression}
    &\left|\GretHTL_{\mu\nu}(P-P')(P+P')^\mu (K+K')^\nu\right|^2 =\frac{c_1^2}{A^2+B^2}+\frac{c_2^2}{C^2+D^2}-\frac{2c_1c_2(AC+BD)}{(A^2+B^2)(C^2+D^2)}\,,
\end{align}
with
\begin{subequations}
    \begin{align}
        c_1 &= (2p+\omega)(2k-\omega)\,, &
        c_2 &= 4pk\sin\theta_{qp}\sin\theta_{qk}\cos(\phi_{qk}-\phi_{qp})\,,\label{eq:constant_c2}\\
        A &= q^2+\RE\Pi^{00}(x)\,, \qquad\qquad B=\IM\Pi^{00}(-x)\,,&
        C &= q^2-\omega^2+\RE\Pi^T(x)\,, \qquad D =\IM\Pi^T(-x)\,.
    \end{align}
\end{subequations}
Here, $\theta_{qp}$ and $\theta_{qk}$ are the angles between $\q$ and $\p$ or $\k$. The angle $\phi_{qk}$ ($\phi_{qp}$) is the azimuthal angle of $\k$ ($\p)$ in a coordinate system where $\q$ points in the $z$ direction and the beam axis lies in the $xz$ plane.

\begin{figure}
    \centering
    \includegraphics[width=0.49\linewidth]{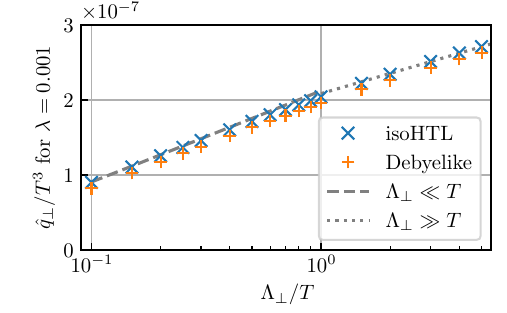}
    \includegraphics[width=0.49\linewidth]{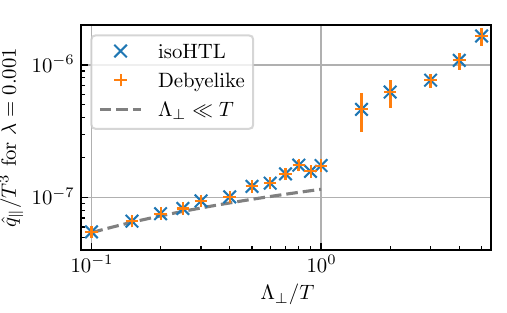}
    \caption{Momentum broadening coefficients for a purely gluonic plasma and gluon jet. The two different screening prescription is compared to known analytic limits. Left: Transverse momentum broadening as quantified by the jet quenching parameter $\hat q_{\perp}$. Right: Longitudinal momentum broadening, $\hat q_\parallel$.}
    \label{fig:momentumbroadening}
\end{figure}
To verify the correct implementation of the HTL screening, we calculate the transverse and longitudinal momentum broadening coefficients $\hat q_\perp$ and $\hat q_\parallel$, for which screening of the soft-gluon exchange in the $t$-channel matrix element is important. For that, we use the implementation \cite{kurkela_2023_10409474} described in Ref.~\cite{Boguslavski:2023waw} for $p\to\infty$ and a transverse momentum cutoff. We compare with their analytic estimates \cite{Aurenche:2002pd, Arnold:2008vd, Ghiglieri:2015ala},
\begin{align}
    \hat q_\perp^{\mathrm{therm}}(\Lambda_\perp\ll T)&=\frac{g^2}{4\pi}C_R T m_D^2\ln\left(1+\frac{\Lambda_\perp^2}{m_D^2}\right)\,,\label{eq:qhat_therm_soft}\\
    \hat q_\perp^{\mathrm{therm}}(\Lambda_\perp\gg T)&=\frac{C_R g^4 T^3}{\pi^2}\sum_{\pm}\Xi_\pm\mathcal I_\pm(\Lambda_\perp)\,,\label{eq:qhat_therm_hard}\\
    \hat q_\parallel^{\mathrm{therm}}(\Lambda_\perp\ll T)&=\frac{g^2C_R T}{2\pi}M_\infty^2\ln\frac{\Lambda_\perp}{M_\infty}\,,\label{eq:qhatL_therm_soft}
\end{align}
where $\mathcal I_\pm(\Lambda_\perp)=\frac{\zeta_\pm(3)}{2\pi}\ln(\Lambda_\perp/m_D)+ \frac{\zeta_\pm(2)-\zeta_\pm(3)}{2\pi}(\ln(T/m_D)+0.5-\gamma_E+\ln 2)-\sigma_\pm/(2\pi)$, where $\zeta_+(x)$ is the Riemann Zeta function and $\zeta_-(s)=(1-2^{1-s})\zeta_+(s)$, $\gamma_E$ is the Euler Mascheroni constant, and $\sigma_\pm=\sum_{k=1}^\infty \frac{(\pm1)^{k-1}}{k^3}\ln[(k-1)!]$. For gluons, $\sigma_+=0.38604\dots$.
In particular, for transverse momentum broadening (left panel in Figure \ref{fig:momentumbroadening}), we compare with the analytic result for small cutoffs \eqref{eq:qhat_therm_soft} (dashed line), and for large cutoffs \eqref{eq:qhat_therm_hard} (dotted line). For longitudinal momentum broadening (right panel), we compare with Eq.~\eqref{eq:qhatL_therm_soft} (dotted line). Note that this analytic result is only valid for small cutoffs, and deviations for larger cutoffs---as seen in the figure---are expected. For both transport coefficients, we show both the Debye-like (orange $+$) and isoHTL screening prescription (blue $\times$), where we use the value $\xi_g=e^{5/6}/\sqrt{8}$ for the Debye-like prescription~\cite{AbraaoYork:2014hbk}. This parameter is tuned to correctly reproduce the isoHTL results for longitudinal momentum broadening, as can be seen in the right panel (resulting from an expansion of the distribution function in $\omega$). However, it fails to reproduce the correct result for the transverse momentum broadening. If we had chosen to use the value $\xi_g'=e^{1/3}/2$ obtained in Ref.~\cite{Boguslavski:2023waw}, we would have found the transverse broadening to be accurately described. We emphasize that the isoHTL screening correctly describes both of these processes and is thus more general and should be used instead \cite{Boguslavski:2024kbd}.

For screening soft-fermionic exchanges, we use 
\cite{Arnold:2002zm}
\begin{align}
    \frac{u}{t}\to \frac{4\text{Re}[(P\cdot \mathcal Q)(K\cdot\mathcal Q)^\ast]+s\mathcal Q\cdot\mathcal Q^\ast}{|\mathcal Q\cdot\mathcal Q|^2}\,,
\end{align}
with $\mathcal Q^\mu = Q^\mu - \Sigma^\mu_{Ret}=(P-P')^\mu - \Sigma^\mu_{Ret}$. For the parameterization of $\mathcal Q$ as 
\begin{align}
    \mathcal Q^\mu = (F_1+iF_2,0,0,F_3+iF_4)\,,
\end{align}
we obtain
\begin{align}
    \frac{u}{t}\to M^f_{HTL}=\frac{4kp\left[A-C\cos\theta_{qp}+\cos\theta_{qk}(-C+B\cos\theta_{qp})\right]+s(B-A)}{((F_1-F_3)^2+(F_2-F_4)^2)((F_1+F_3)^2+(F_2+F_4)^2)}\,,\label{eq:fermionic-screening-prescription}
\end{align}
where
\begin{align}
    A &= F_1^2+F_2^2\,,& B&= F_3^2+F_4^2\,,&
    C&= F_1F_3+F_2F_4\,.
\end{align}
We used again the same integration variables that are described below \ref{eq:constant_c2}.
For $s/u$, appearing in the $qg\leftrightarrow qg$ matrix element, the replacement is slightly different and does not amount to simply relabeling $u\leftrightarrow t$, but the differences are negligible for small $|u|\ll s$ and $|\mathcal Q\cdot \mathcal Q|^2\ll  s\mathcal Q\cdot \mathcal Q^\ast$, where the screening is needed.

In practice, we replace $(s-u)/t$, which can be related to
\begin{align*}
    \frac{u}{t}=\frac{1}{2}\left(\frac{u}{t}+\frac{-s-t}{t}\right)=\frac{1}{2}\left(\frac{u-s}{t}-1\right),
\end{align*}
thus,
\begin{align}
    \frac{s-u}{t}&=-\frac{2u}{t}-1\to -2M^f_{\mathrm{HTL}}-1.
\end{align}
We work in a frame in which $Q^\mu=(\omega,0,0,q)$, and using the explicit form of the fermionic self-energy \cite{Arnold:2002zm} 
\begin{align}
    \Sigma^0(Q)&=\frac{m^2}{4q}\left(\ln\left|\frac{\omega+q}{\omega -q}\right|-i\pi\right),&
    \Sigma^3(Q)&=-\frac{m^2}{2q}\left\{1-\frac{\omega}{2q}\left[\ln\left|\frac{\omega+q}{\omega -q}\right|-i\pi\right]\right\}\,,
\end{align}
we find the explicit forms of the $F_i$,
\begin{align}
    F_1 &= \omega - \frac{m^2}{4q}L\,, & F_2 &= \frac{m^2\pi}{4q}\,,\\
    F_3 &= q+\frac{m^2}{2q}\left(1-\frac{\omega}{2q}L\right)\,, & F_4 &= \frac{m^2\omega}{4q^2}\pi=F_2\frac{\omega}{q}\,,
\end{align}
where $L=\ln\left|\frac{\omega+q}{\omega -q}\right|$, and 
$m \equiv m_{\mathrm{eff}}^f$ is the gluonic asymptotic (or effective) mass. While our definition of $Q=P'-P$ differs from the one in \cite{Arnold:2002zm} by $Q^\mu\to -Q^\mu$, this does not change the implementation of the screening prescription. It would reverse $\omega\to -\omega$ and $\Sigma^3\to-\Sigma^3$. The latter implies a sign change in $F_4$ and of the parenthesis in $F_3$. The $\omega$ change is straightforward and amounts to $\omega\to-\omega$. Also note that $L\to -L$. Overall, this results in a sign change in $F_1$ and $F_3$, which only appear in pairs or squared, thus not affecting the result in Eq.~\eqref{eq:fermionic-screening-prescription}.

\begin{figure}
    \centering
    \includegraphics[width=0.49\linewidth]{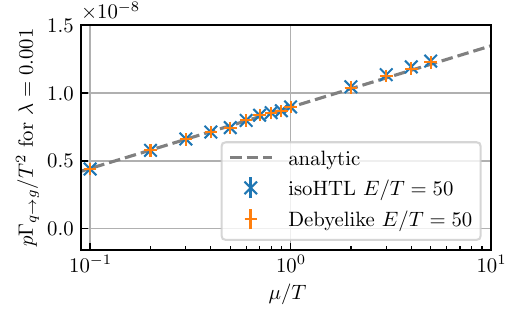}
    \caption{Numerical evaluation of the quark to gluon conversion rate with Debye-like ($+$) and HTL ($\times$) screened matrix elements.
    The analytic result \eqref{eq:conversion-rate-analytic} is shown as a gray dashed line. 
    }
    \label{fig:conversionrate}
\end{figure}

For verifying the fermionic HTL implementation, we compare our numerical results of the conversion rate to the expression~\cite{Ghiglieri:2015ala},\footnote{Note that there is a square missing in the final result of the reference, but it can be easily obtained using the previous step.} 
\begin{align}
    \Gamma^{\mathrm{conv}}_{q\to g}(p)=\frac{g^2 C_F (m_{\mathrm{eff}}^f)^2}{8\pi p}\ln\left(1+\frac{\mu^2}{(m_{\mathrm{eff}}^f)^2}\right)\,,\label{eq:conversion-rate-analytic}
\end{align}
where $\mu$ is an infrared (or UV) cutoff used in the $q_\perp$ integration which separates the hard from the soft sector in the reorganization of the scattering processes introduced in Ref.~\cite{Ghiglieri:2015ala}. Consequently, this formula is only valid for large jet energies and small cutoffs $\mu$ and our implementation (for finite $p$) is only valid for large jet energies $p\gg T$. The results of this comparison are shown in Figure \ref{fig:conversionrate}, with the $+$ signs from the Debye-like screening prescription and the crosses showing the HTL prescription. It can be seen that the Debye-like screening agrees well with the isoHTL screening. This is because this process was used to fix the parameter $\xi_q$ in the Debye-like screening prescription~\cite{Ghiglieri:2015ala,Kurkela:2018oqw}. While in Fig.~\ref{fig:conversionrate}, we use $p=50T$, we have checked that $p\geq 10$ leads to the same numerical results, when rescaled with $p$.

\section{Expectation values in the single collision limit}
\label{app:Expectation_values_OE}
The expectation value of a general observable $\mathcal O(\p,i)$ is
\begin{equation}
    \begin{split}
        \hat{\mathcal O}(t) &= \sum_i\nu_i\int_{\p}\mathcal O(\p,i) \delta f_i(\p,t)\,,
    \end{split}
\end{equation}
where $f$ is the solution of the Boltzmann equation. Note that, relative to \cref{eq:exp_value}, one has to divide by the initial density, which is $\nu_{i_0} \epsilon\, T^3$ for a delta perturbation. We saw examples for $\mathcal O$ in the main text and here we are focusing on its evaluation in the single collision limit. Following Eq.~\eqref{eq:single-hit}, its rate takes the form
\begin{equation}
    \partial_t \hat{\mathcal O}=\sum_i\nu_i\int_{\p}\mathcal O(\p,i) \delta C_i[\{n,\delta f^{(0)}\}]\,.
\end{equation}
The contribution from the elastic collisions is given by
\begin{equation}
    \begin{split}
        \partial_t\hat{\mathcal O}
        &\overset{2\leftrightarrow2}{=}\frac{1}{2}\sum_{abcd}\int_{p}\int d\Omega^{2\leftrightarrow2}\mathcal O(\p,a)|\mathcal M^{ab}_{cd}(\p,\k;\p',\k')|^2\delta\mathcal F(\p,\k;\p',\k')\,.
    \end{split}
\end{equation}
By integrating the initial perturbation $\delta f_i^{(0)}(\p)=\delta_{ii_0}(2\pi)^3\delta^3(\p-\p_0)$, relabeling sums and integrals, and using matrix element symmetries
\begin{equation}
    |\mathcal M^{ab}_{cd}(\p,\k;\p',\k')|^2=|\mathcal M^{ba}_{cd}(\k,\p;\p',\k')|^2=|\mathcal M_{ab}^{cd}(\p',\k';\p,\k)|^2=|\mathcal M^{ab}_{dc}(\p,\k;\k',\p')|^2\,,
\end{equation}
we arrive to the following simple formula
\begin{equation}
    \begin{split}
        \partial_t\hat{\mathcal O}
        &\overset{2\leftrightarrow2}{=} \frac{1}{4p_0}\sum_{bcd}\int d\Omega^{2\leftrightarrow2}|\mathcal{M}^{i_0b}_{cd}(\p_0,\k;\p',\k')|^2 \\
        &\quad\times \left\{n_b(k)(1\pm n_c(p'))(1\pm n_d(k'))\mp_{i_0} n_c(p')n_d(k')(1\pm n_b(k))\right\}\\
        &\quad\times \left[ \mathcal O(\p',c)+\mathcal O(\k',d) -\mathcal O(\p_0,i_0) - \mathcal O(\k,b) \right] \,.
    \end{split}\label{eq:c22-genral-O-final}
\end{equation}
The contribution from the radiation process is
\begin{equation}
    \begin{split}
        \partial_t\hat{\mathcal O}
        &\overset{1\leftrightarrow2}{=} \sum_a\int_{\p}\mathcal O(\p,a)\frac{(2\pi)^3}{2p^2}\sum_{bc}\int_0^\infty dp'dk'\delta(p-p'-k')\gamma^{a}_{bc}(\p;p'\hat\p,k'\hat\p)\delta\mathcal F(\p;p'\hat\p,k'\hat\p)\\
        &-\sum_a\int_{\p}\mathcal O(\p,a)\frac{(2\pi)^3}{p^2}\sum_{bc}\int_0^\infty dp'dk'\delta(p+k'-p')\gamma^c_{ab}(p'\hat\p;\p,k'\hat\p)\delta\mathcal F(p'\hat\p;\p,k'\hat\p)\,.
    \end{split}
\end{equation}
Similar to the elastic collisions, by integrating the perturbation $\delta f_i(t_0,\p)=(2\pi)^3\delta_{ii_0}\frac{1}{p^2}\delta(p-p_0)\delta^2(\hat\p-\hat\p_0)$, relabeling variables, and using symmetries $\gamma^a_{bc}(\p;\p',\k')=\gamma^a_{cb}(\p;\k',\p')$,
we arrive at the formula
\begin{equation}
    \begin{split}
        \partial_t\hat{\mathcal O}
        &\overset{1\leftrightarrow2}{=} \frac{(2\pi)^3}{p_0^2}\sum_{ac}\int dp dk'\left[\mathcal O(p\hat\p_0,a)-\mathcal O(\p_0,i_0)-\mathcal O(k'\hat\p_0,c)\right]\delta(p-p_0-k')\gamma^{a}_{i_0c}(p\hat\p_0;p_0\hat\p_0,k'\hat\p_0)\\
        &\quad\times\left[n_c(k')(1\pm n_a(p))\mp_{i_0} n_a(p)(1\pm n_c(k'))\right]\\
        &+\frac{(2\pi)^3}{2p_0^2}\sum_{ab}\int dpdk\left[\mathcal O(p\hat\p_0,a)+\mathcal O(k\hat\p_0,b)-\mathcal O(\p_0,i_0)\right]\delta(p+k-p_0)\gamma^{i_0}_{ab}(p_0\hat\p_0;p\hat\p_0,k\hat\p_0)\\
        &\quad\times\left[(1\pm n_a(p))(1\pm n_b(k))\mp_{i_0} n_a(p)n_b(k)\right] \,.
    \end{split}
\end{equation}

Enforcing $p>k$ in the second term and integrating over $p$, we obtain
\begin{equation}
    \begin{split}
        \partial_t\hat{\mathcal O}
        &\overset{1\leftrightarrow2}{=} \frac{(2\pi)^3}{p_0^2}\sum_{ac}\int_0^\infty dk'\left[\mathcal O((p_0+k')\hat\p_0,a)-\mathcal O(\p_0,i_0)-\mathcal O(k'\hat\p_0,c)\right]\gamma^{a}_{i_0c}((p_0+k')\hat\p_0;p_0\hat\p_0,k'\hat\p_0)\\
        &\quad\times\left[n_c(k')(1\pm n_a(p_0+k'))\mp_{i_0} n_a(p_0+k')(1\pm n_c(k'))\right]\\
        &+\frac{(2\pi)^3}{p_0^2}\sum_{ab}\int_0^{p_0/2} dk\left[\mathcal O((p_0-k)\hat\p_0,a)+\mathcal O(k\hat\p_0,b)-\mathcal O(\p_0,i_0)\right]\gamma^{i_0}_{ab}(p_0\hat\p_0;(p_0-k)\hat\p_0,k\hat\p_0)\\
        &\quad\times\left[(1\pm n_a(p_0-k))(1\pm n_b(k))\mp_{i_0} n_a(p_0-k)n_b(k)\right] \,.
    \end{split}\label{eq:c12-general-final}
\end{equation}
The final answer is the sum of the contributions $2\leftrightarrow2$ and $1\leftrightarrow2$. These formula agrees with previous studies~\cite{Arnold:2000dr,Arnold:2003zc}.

\section{Numerical implementation of the single-collision limit\label{eq:numerics}}
Here, we discuss how the integral \eqref{eq:c22-genral-O-final} is performed. The integral measure can be rewritten similarly as in Ref.~\cite{Boguslavski:2023waw},
\begin{equation}
    \begin{split}
        \int\dd\Omega^{2\leftrightarrow 2} &= \int_{p_2 p_3 p_4}(2\pi)^4\delta^4(P+P_2-P_3-P_4)\\
        &=\frac{1}{4p(2\pi)^5}\int_0^{2\pi}\dd{\phi_{pq}}\int_0^{2\pi}\dd{\phi_{kq}}\int_0^\infty\dd{k}\int_{-\frac{p-k}{2}}^k\dd{\omega}\int_{|\omega|}^{\min(2p+\omega, 2k-\omega)}\dd{q}\,.
    \end{split}
\end{equation}
We have used the symmetry $p'\leftrightarrow k'$ in the integrand (also present in \eqref{eq:c22-genral-O-final}) to always enforce $p'>k'$. Otherwise, the lower boundary of the $\omega$-integral would be $-p$, and the prefactor $1/(8p(2\pi)^5))$. The reason for this choice is that then only the $t$ channel diagrams need to be screened \cite{Boguslavski:2024kbd}. For a finite momentum grid $p_{\mathrm{min}}<p<p_{\mathrm{max}}$, on which the distribution function is stored, we require $k$ to be on the grid, and $k'>p_{\mathrm{min}}$,
which results in
\begin{align}
    \int\dd\Omega^{2\leftrightarrow 2}=\frac{1}{4p(2\pi)^5}\int_0^{2\pi}\dd{\phi_{pq}}\int_0^{2\pi}\dd{\phi_{kq}}\int_{\max(p_{\mathrm{min}},2p_{\mathrm{min}}-p)}^{p_{\mathrm{max}}}\dd{k}\int_{-\frac{p-k}{2}}^{k-p_{\mathrm{min}}}\dd{\omega}\int_{|\omega|}^{\min(2p+\omega, 2k-\omega)}\dd{q}\,.
\end{align}
This integration measure is used in Eq.~\eqref{eq:c22-genral-O-final} together with the matrix elements tabulated in Ref.~\cite{Boguslavski:2023waw}, with screening prescriptions \eqref{eq:HTL_propagator_explicit_expression} and \eqref{eq:fermionic-screening-prescription} to obtain the expectation values in the single collision limit. Similar as in Ref.~\cite{Boguslavski:2023waw}, the five-dimensional integral is evaluated using Monte Carlo integration with importance sampling.
The kinematic variables need to be expressed in terms of the integration variables. The Mandelstam variables are given by
\begin{align}
    t=\omega^2-q^2\,, && u=\frac{t}{2q^2}\left((p+p')(k+k')-q^2-\sqrt{(4pp'+t)(4k'k+t)}\cos\phi_{kq}\right)\,, && s=-t-u\,,
\end{align}
and the relevant angles are
\begin{align}
    \cos\theta_{pq}=\frac{\omega}{q}+\frac{t}{2pq}\,, && \cos\theta_{kq}=\frac{\omega}{q}-\frac{t}{2kq}\,, && \cos\theta_{k'q}=\frac{\omega}{q}+\frac{t}{2k'q}\,,
\end{align}
where $k'=k-\omega$, and $p'=p+\omega$. We also need to express $\p=(0,0,p)$, $\k$, $\k'$, and $\p'$ in the frame defined by the direction of propagation of the jet,
\begin{align}
    \k&= k \begin{pmatrix}
        \cos\phi_{pq}(\cos\phi_{kq}\cos\theta_{pq}\sin\theta_{kq} + \cos\theta_{kq}\sin\theta_{pq})-\sin\phi_{pq}\sin\phi_{kq}\sin\theta_{kq}\\
        \cos\phi_{pq}\sin\phi_{kq}\sin\theta_{kq}+\sin\phi_{pq}\left(\cos\phi_{kq}\cos\theta_{pq}\sin\theta_{kq}+\cos\theta_{kq}\sin\theta_{pq}\right)\\
        \cos\theta_{kq}\cos\theta_{pq}-\cos\phi_{kq}\sin\theta_{kq}\sin\theta_{pq}
    \end{pmatrix}, && \q=q\begin{pmatrix}
        \sin\theta_{pq}\cos\phi_{pq}\\\sin\theta_{pq}\sin\phi_{pq}\\\cos\theta_{pq}
    \end{pmatrix}, 
\end{align}
and $\k'=\k-\q$, and $ \p'=\p+\q$. In our implementation of Eq.~\eqref{eq:c22-genral-O-final}, we included both the gain and loss term and verified that for all observables considered in this paper, the gain term is negligible.

To compare with the analytic expression for the conversion rate \eqref{eq:conversion-rate-analytic}, we need to additionally implement a momentum cutoff on $-t=q^2-\omega^2<\mu^2$ \cite{Ghiglieri:2015ala}. This changes the upper boundary of the $q$-integral to $\min(2p+\omega, 2k-\omega, \sqrt{\omega^2+\mu^2})$.

For the inelastic collision term \eqref{eq:c12-general-final}, the kinematics is simpler. Getting rid of the delta function there, only one integral needs to be performed numerically, for which we use both the trapezoidal and Simpson's rule, and use their difference as an error estimate. Compared to simulating the time evolution of minijets \cite{Zhou:2024ysb}, the evaluation of these integrals is fast and the error can be easily reduced by a finer momentum grid.

\section{Numerical convergence of collision kernel evaluation in EKT\label{app:convergence}}
In this section of the Appendix, we discuss the numerical setup needed for the evolution of a minijet from which we compute the transport coefficients in \cref{sec:resultstransport}. We will demonstrate numerical convergence for the transport coefficient $\hat{q}_{\perp}(\Lambda_{\textrm{min}})$ (\cref{eq:qhat_cutoff}). Computing the transport coefficients from the minijet evolution, using \cref{eq:qhat_cutoff} evaluated after one time step, is numerically challenging due to the fact that the initial condition is representing a delta-like perturbation.
We use spherical coordinates to discretize our grid in momentum space, so we consider the magnitude $p$, the polar angle $v_z = \cos\theta$ and the azimuthal angle $\phi$. In all our computations we assume azimuthal symmetry. Importantly, we have $p_z = pv_z$ and $p_{\perp} = p\sqrt{1-v_z^2}$. For the $p$-discretization we use a linear grid with $N_p$ grid points with grid spacing $\Delta p = (p_{max} - p_{min})/N_p$. For the angle $v_z = \cos\theta$ we use a logarithmic grid with $N_{v_z}$ grid points. Namely, we discretize $w(v_z) = \log(1+v_z^0 - v_z)$ linearly, where $v_z^0$ is a small offset so that the logarithm is well defined. We choose $v_z^0 = 10^{-7}$. The number of grid points for the angle is denoted by $N_{v_z}$.

In order to solve for the time evolution of the minijet in \cref{eq:LB}, we have to calculate the linearized collision kernel on the right-hand side. For one time step, this is equivalent to computing \cref{eq:single-hit}. For the evaluation of the multidimensional integral, we use Monte-Carlo integration. The total number of samples $N_{\textrm{samples}}$ chosen to execute the integration differs between the elastic kernel in \cref{eq:C22} and the inelastic one in \cref{eq:C12}. For the inelastic kernel we choose $N_{\textrm{samples}}^{1\leftrightarrow2} = K\times N_p$ whereas for the elastic one we have $N_{\textrm{samples}}^{2\leftrightarrow2} = K \times N_p\times N_{v_z}$, where $K$ is some large factor. This is because in our leading order framework, the splittings are strictly collinear, and we need to sample over a smaller phase space containing only the momentum $p$ and not the angle $v_z$. For minijet results in \cref{sec:resultstransport} (bands in \cref{fig:qhat_cutoff_lambda,fig:qhat_L_sub_cutoff_lambda,fig:qhat_L_cutoff_lambda,fig:drag_cutoff_lambda}) we choose $N_p = 1000$ and $N_{v_z} = 300$ and the sampling multiplier $K  =  2\cdot 10^4$.
\begin{figure}
    \centering
    \includegraphics[width=0.8\linewidth]{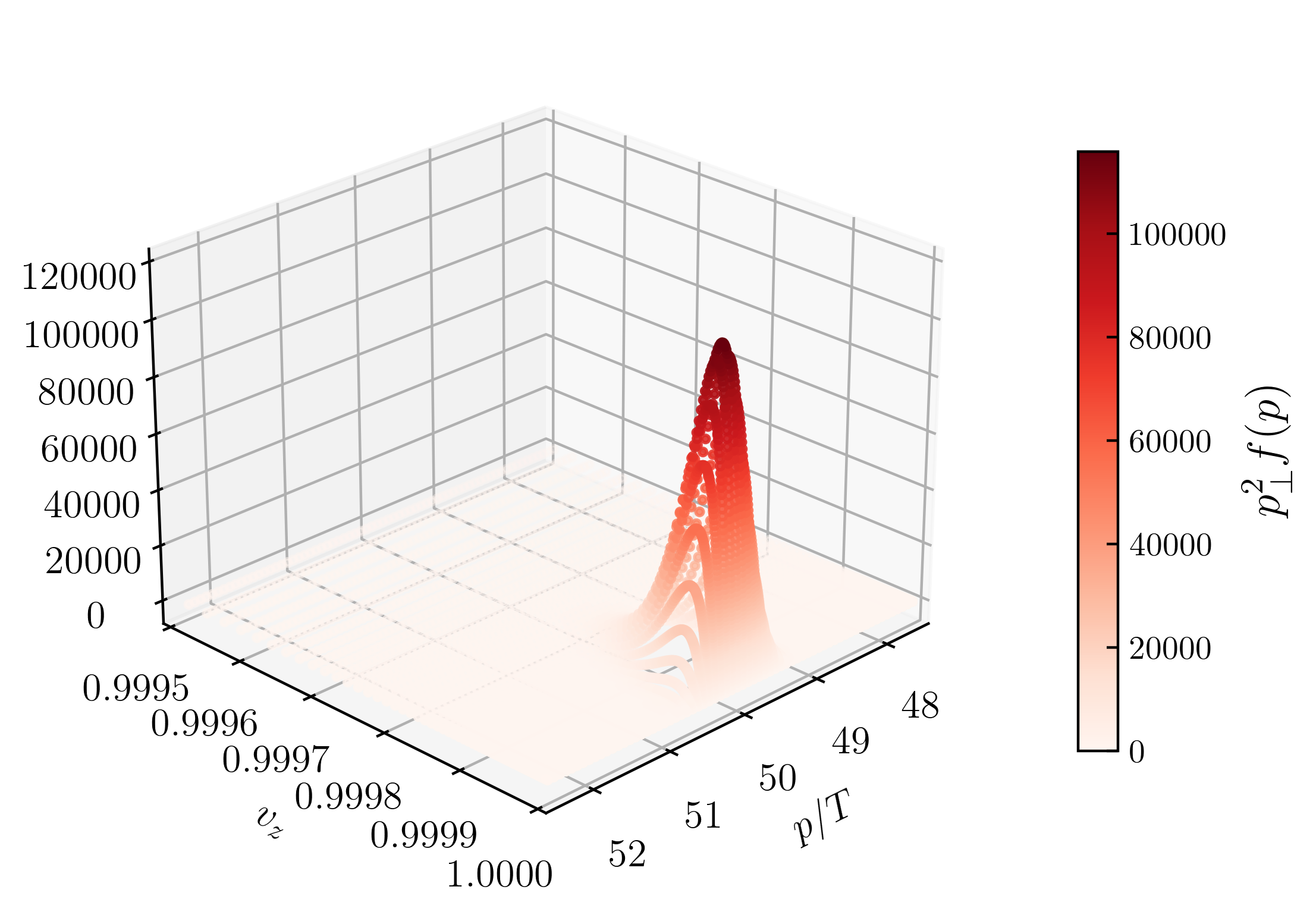}
    \caption{Initial condition of the jet perturbation from \cref{eq:perturbation} weighted by $p_{\perp}^2$ , with $\sigma = 0.005E$. The momentum range plotted corresponds to $\sim 10 \sigma$.}
    \label{fig:3d_IC}
\end{figure}
\begin{figure}
    \centering
    \includegraphics[width=1\linewidth]{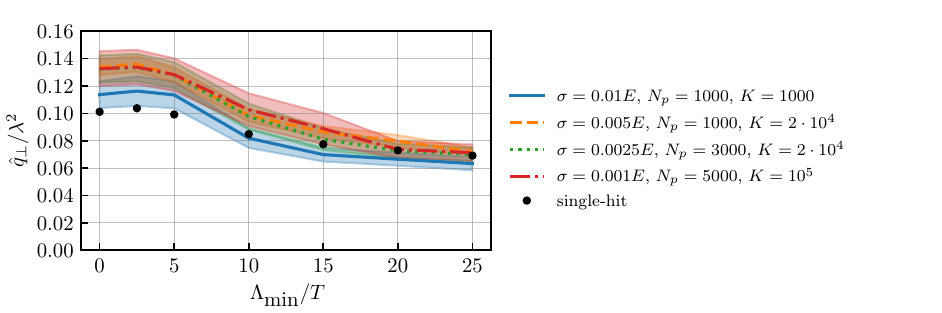}
    \caption{Transverse momentum broadening coefficient $\hat{q}_{\perp}$ for initial minijet energy $E=50T$ and $\lambda=10$ as a function of lower momentum cutoff $p>\Lambda_\text{min}$. Different colors correspond to different widths $\sigma$ using different discretizations. Bands indicate statistical uncertainty of a single-step evolution of a gaussian perturbation in EKT, while points show the direct Monte-Carlo evaluation of a single-hit transport integral \cref{eq:qhat_cutoff}.}
    \label{fig:different_sigma_with_error}
\end{figure}

In order to show that $\hat{q}_{\perp}$ has reached numerical convergence, we consider various widths $\sigma$ of the initial Gaussian. Sharper peaks require finer grids which amounts to increasing $N_p$. For large $N_p$, the grid spacing scales $\Delta p/T \sim 1/N_p$. For each setting we ensure that $1/N_p \ll \sigma/T$. For the angle, we choose a logarithmic grid to ensure that the angular discretization around the initial peak is very fine, i.e., the angular grid spacing goes $\Delta v_z\sim v_z^0/N_{v_z}\ll \sigma/T$ (cf. \cref{fig:3d_IC}). For this reason, we will use the same $v_z$-discretization of $N_{v_z} = 300$ for all of the curves.

We compute $\hat{q}_{\perp}(\Lambda_{\textrm{min}})$ for four different widths $\sigma$, successively increasing $N_p$ and $K$. This is shown in \cref{fig:different_sigma_with_error}. The first curve for $\sigma = 0.01E$ (blue) lies below the rest of the curves. By going to smaller $\sigma$, the value of $\hat{q}_{\perp}(\Lambda_{\textrm{min}})$ approaches the one we show in \cref{fig:qhat_cutoff_lambda} for $\lambda=10$. In particular, this holds for all $\Lambda_{\textrm{min}}$, with the same conclusion that the single-step evolution results are systematically larger than the single-hit values at low cutoffs. So in \cref{fig:qhat_cutoff_lambda,fig:qhat_L_sub_cutoff_lambda,fig:qhat_L_cutoff_lambda,fig:drag_cutoff_lambda}, the default width we are using is $\sigma = 0.005E$.

In \cref{sec:resultsthermalization}, we consider long time EKT evolution for perturbations with $\sigma=0.01E$, until the perturbation reaches equilibrium. In this case, we used a different grid  discretization. To have good resolution for small momenta  $p\lesssim T$ we used logarithmic discretization in $p$. At late times the distribution approaches isotropy, therefore we used linearized discretization in $v_z$. 
The grid parameters for these simulations are $N_p = 150$, $N_{v_z} = 150$ and $K = 120$. Although the $K$ factor is much smaller than in a single-step simulations above, the average over many successive time-steps partially compensate for less accurate kernel integrals at each time-step.

\end{document}